\documentclass[sigconf]{acmart}

\AtBeginDocument{%
  \providecommand\BibTeX{{%
    \normalfont B\kern-0.5em{\scshape i\kern-0.25em b}\kern-0.8em\TeX}}}

\usepackage{amsmath,amsfonts}
\usepackage{algorithmic}
\usepackage{graphicx}
\usepackage{textcomp}
\usepackage{xcolor}
\usepackage{amsthm}
\usepackage{makecell}
\usepackage{float}
\usepackage{subfig}
\usepackage{graphicx}
\usepackage{xspace}
\usepackage{url}
\usepackage{hyperref}
\usepackage{multirow}

\usepackage{pgfplots}
\usepackage{numprint}
\usepackage{nicefrac}
\usepackage{amsthm}
\newtheorem{definition}{Definition}

\usepackage{array}
\newcolumntype{L}[1]{>{\raggedright\let\newline\\\arraybackslash\hspace{0pt}}m{#1}}
\newcolumntype{C}[1]{>{\centering\let\newline\\\arraybackslash\hspace{0pt}}m{#1}}

\def\BibTeX{{\rm B\kern-.05em{\sc i\kern-.025em b}\kern-.08em
    T\kern-.1667em\lower.7ex\hbox{E}\kern-.125emX}}
    
\usepackage{array}
\newcolumntype{?}{!{\vrule width 1pt}}

 \setcopyright{none} 

\renewcommand\footnotetextcopyrightpermission[1]{} 

\begin{document}

\newcommand{\dquotes}[1]{``#1''}
\newcommand{\squotes}[1]{`#1'}

\newcommand{\algcomment}[1]{%
    \vspace{-\baselineskip}%
    \noindent%
    {\footnotesize #1\par}%
    \vspace{\baselineskip}%
    }
\definecolor{ao(english)}{rgb}{0.0, 0.5, 0.0}

\def\amazonmen{\texttt{Amazon Men}\xspace}
\def\amazonwomen{\texttt{Amazon Women}\xspace}
\def\tradesy{\texttt{Tradesy}\xspace}

\newcommand{\daniele}[1]{\textcolor{magenta}{{\bf [Daniele: }{ \em #1}{\bf ]}}}
\newcommand{\felice}[1]{\textcolor{blue}{{\bf [Felice: }{ \em #1}{\bf ]}}}
\newcommand{\walter}[1]{\textcolor{red}{{\bf [Walter: }{ \em #1}{\bf ]}}}

\newcommand{\dm}[1]{\textcolor{magenta}{{ #1}}}
\newcommand{\flc}[1]{\textcolor{blue}{{ #1}}}
\newcommand{\wa}[1]{\textcolor{red}{{ #1}}}
\newcommand{\dsml}[1]{\textcolor{ao(english)}{{ #1}}}

\def\recprob{\textit{Recommendation Problem}\xspace}
\def\taamr{\textit{TAaMR}\xspace}
\def\amazonmen{\texttt{Amazon Men}\xspace}
\def\amazonwomen{\texttt{Amazon Women}\xspace}
\def\tradesy{\texttt{Tradesy}\xspace}
\def\resnet{\texttt{ResNet50}\xspace}
\def\imagenet{\texttt{ImageNet}\xspace}
\def\categoryhitratio{Category Hit Ratio\xspace}
\def\categoryhitratioat{Category Hit Ratio@N\xspace}
\def\chr{$CHR@K$\xspace}
\def\chratventi{$CHR@20$\xspace}
\def\chratcinq{$CHR@50$\xspace}
\def\chratcento{$CHR@100$\xspace}
\def\cndcg{$nCDCG@K$\xspace}
\def\cndcgatventi{$nCDCG@20$\xspace}
\def\cndcgatcinq{$nCDCG@50$\xspace}
\def\cndcgatcento{$nCDCG@100$\xspace}

\def\var{\texttt{VAR}\xspace}

\author{Vito Walter Anelli, Tommaso Di Noia, Daniele Malitesta,  Felice Antonio Merra}
\email{{vitowalter.anelli, tommaso.dinoia, daniele.malitesta, felice.merra}@poliba.it}
\affiliation{%
  \institution{Polytechnic University of Bari}
  \city{Bari}
  \state{Italy}
}
\authornote{The authors are in alphabetical order. Corresponding author: Felice Antonio Merra (\texttt{felice.merra@poliba.it}), Daniele Malitesta (\textit{daniele.malitesta@poliba.it}).}

\title{An Empirical Study of DNNs Robustification Inefficacy in Protecting Visual Recommenders}

\renewcommand{\shortauthors}{Anelli, Di Noia, Malitesta, and Merra}

\begin{abstract}
Visual-based recommender systems (VRSs) enhance recommendation performance by integrating users' feedback  with the visual features of product images extracted from a deep neural network (DNN).
Recently, human-imperceptible images perturbations, defined \textit{adversarial attacks}, have been demonstrated to alter the VRSs recommendation performance, e.g., pushing/nuking category of products.
However, since adversarial training techniques have proven to successfully robustify DNNs in preserving classification accuracy, to the best of our knowledge, two important questions have not been investigated yet: 1) How well can these defensive mechanisms protect the VRSs performance? 2) What are the reasons behind ineffective/effective defenses?
To answer these questions, we define a set of defense and attack settings, as well as recommender models, to empirically investigate the efficacy of defensive mechanisms. The results indicate alarming risks in protecting a VRS through the DNN robustification. Our experiments shed light on the importance of visual features in very effective attack scenarios. Given the financial impact of VRSs on many companies, we believe this work might rise the need to investigate how to successfully protect visual-based recommenders.
Source code and data are available at \url{https://anonymous.4open.science/r/868f87ca-c8a4-41ba-9af9-20c41de33029/}.

\end{abstract}


\maketitle

\section{Introduction}\label{sec:introduction}
Recommender Systems (RSs) have terrifically taken over online shopping by providing personalized recommendations to users in the flood of products of e-commerce platforms. Catching a large number of historical interactions, RSs learn what each user might like, and show short ranked lists of the presumably desired products. In domains such as fashion, food, and point-of-interest recommendations, images are associated with items to get customers' attention. Visual-based Recommender Systems (VRSs) are the cornerstone of recommender models to learn users' preferences by mixing past interactions with high-level visual features extracted from those item photos.
The intuition behind this class of recommenders is that users' preference is influenced by the observable style of product images. 
Thanks to the power of Deep Neural Networks (DNNs) in capturing high-level visual aspects, the state-of-the-art of VRSs incorporates deep visual features extracted from Convolutional Neural Networks (CNNs). For instance, He \textit{et al.}~\cite{DBLP:conf/aaai/HeM16, DBLP:conf/www/HeM16} proposed one of the first models, named VBPR, to integrate visual features getting outperforming recommendation performance compared to the basic version of the recommender (BPR-MF ~\cite{DBLP:conf/uai/RendleFGS09}). 


While different variants of VRSs have been proposed in the last years ---taking the quality of DNNs-extracted features for granted --- the work by Szegedy \textit{et al.}~\cite{DBLP:journals/corr/SzegedyZSBEGF13} raised security concerns as they showed that a malicious person, an \textit{adversary}, may lead the network to misclassify an image corrupted by a human-imperceptible \textit{adversarial perturbation}.
Starting from Szegedy's publication, different adversarial strategies (e.g., FGSM~\cite{DBLP:journals/corr/GoodfellowSS14}, PGD~\cite{DBLP:conf/iclr/MadryMSTV18}, and Carlini \& Wagner~\cite{DBLP:conf/sp/Carlini017}) have demonstrated stronger and stronger attack power. In parallel, a complementary branch of research has been devoted to building robust DNNs (e.g., Adversarial Training~\cite{DBLP:journals/corr/GoodfellowSS14}, Free Adversarial Training~\cite{DBLP:conf/nips/ShafahiNG0DSDTG19}). Consequently, the term \textit{Adversarial Machine Learning} (AML) currently denotes the study of such attacks and defenses.

Motivated by the attacks abilities and the partial protection of the state-of-the-art defense strategies, we identify VRSs as the category of RSs most at risk. Indeed, while AML techniques have been widely investigated in various domains (e.g.,  object detection~\cite{DBLP:conf/nips/RenHGS15}, malware detection~\cite{DBLP:conf/sigcomm/YuanLWX14}, speech recognition~\cite{speech}, graph~\cite{DBLP:conf/wsdm/EntezariADP20}), studies in recommendation scenarios have been conducted only recently. For instance, He \textit{et al.}~\cite{DBLP:conf/sigir/0001HDC18} demonstrated the weakness of matrix factorization recommenders with respect to adversarial perturbations on model embeddings and proposed an adversarial training procedure to make the system robust. Similarly, Tang \textit{et al.}~\cite{8618394} verified the efficacy of adversarial training in protecting VBPR (i.e., a VRS) from perturbations on image features. Moreover, Di Noia \textit{et al.}~\cite{DSML20} proved that targeted adversarial attacks, i.e., FGSM and PGD, applied directly to input images (and not their features) can disturb the recommendation performance.

Differently from the previous works, in this work we propose an empirical framework, \textit{Visual Adversarial Recommendation} (\var), to investigate whether state-of-the-art defense strategies applied to robustify the Image Feature Extractor (IFE) component of a VRS are capable to mitigate the effect of up-to-date adversarial attack strategies, i.e., FGSM, PGD, and even Carlini \& Wagner~\cite{DBLP:conf/ccs/Carlini017, DBLP:conf/icml/AthalyeC018}. 
The motivational scenario involves a competitor willing to increase the recommendability of a category of products on an e-commerce platform (e.g., sandals) by simply uploading adversarially perturbed product images that are misclassified by the IFE as a much more popular class (e.g., running shoes).

The main contributions of this work are summarized as follows:
\begin{itemize}
    \item Study the efficacy of IFE defense approaches in protecting the recommender through the analysis of $54$ combinations of defenses, attacks, and recommendation approached on three real-world datasets.
    \item Joint evaluation of the alteration  of visual recommendation and features extraction performance, with a particular focus on the variation of feature loss on perturbed images.
    \item Propose a novel rank-based evaluation metric, named Category normalized Discounted Cumulative Gain, to deeply explore the efficacy of defenses (or the effects of attacks).
    \item Analyze the variation of global and beyond-accuracy recommendation performance with (and without) defenses applied under the most powerful attack scenarios.
\end{itemize}
The rest of the paper is organized as follows. First, we review related work in Section~\ref{sec:related}. Then, we present the experimental framework in Section~\ref{sec:proposed_framewok}. In Sections~\ref{sec:experiment} and~\ref{sec:discussion} we introduce the experimental setups and present and discuss the empirical results. Finally, we draw conclusion and raise open directions in Section~\ref{sec:conclusion}.


\section{Related Work}\label{sec:related}
\textbf{Adversarial Machine Learning.}
ML models have demonstrated vulnerabilities to adversarial attacks~\cite{DBLP:journals/corr/SzegedyZSBEGF13, DBLP:conf/pkdd/BiggioCMNSLGR13}, i.e., specifically created data samples able to mislead the model despite being highly similar to their clean version. 
Particularly, great research effort has been put into finding the minimum \textit{visual perturbation} to attack images to fool CNN classifiers. Szegedy \textit{et al.}~\cite{DBLP:journals/corr/SzegedyZSBEGF13} formalized the adversarial generation problem by solving a box-constrained L-BFGS. Goodfellow \textit{et al.} proposed Fast Gradient Sign Method (FGSM)~\cite{DBLP:journals/corr/GoodfellowSS14}, a simple one-shot attack method that uses the sign of the gradient of the loss function. Basic Iterative Method (BIM)~\cite{DBLP:conf/iclr/KurakinGB17a} and Projected Gradient Descent (PGD)~\cite{DBLP:conf/iclr/MadryMSTV18} re-adapted FGSM to create stronger attacks by \textit{iteratively} updating the adversarial perturbation.
Carlini and Wagner~\cite{DBLP:conf/sp/Carlini017} improved the problem definition presented in~\cite{DBLP:journals/corr/SzegedyZSBEGF13} and built attacks powerful in deceiving several detection strategies~\cite{DBLP:conf/ccs/Carlini017}. 
Along with the proposed attacks, many solutions have also been provided regarding defense. Adversarial Training~\cite{DBLP:journals/corr/GoodfellowSS14} creates new adversarial samples at training time, making the model more robust to such perturbed inputs. Defensive Distillation~\cite{DBLP:conf/sp/PapernotM0JS16} transfers knowledge between two networks to reduce the resilience to adversarial samples, but was proven not to be as secure as expected against C \& W attacks~\cite{DBLP:journals/corr/CarliniW16}. Free Adversarial Training~\cite{DBLP:conf/nips/ShafahiNG0DSDTG19} truly eases the computational complexity of standard adversarial training without giving up its effectiveness.

\textbf{Security of Visual-based Recommender Systems.}
In this work, the recommendation component is a visual-based recommender model. Different works have demonstrated that the integration of image features in user's preference predictor leads to outperforming both recommendation~\cite{DBLP:conf/www/HeM16, DBLP:conf/aaai/HeM16, DBLP:conf/wsdm/NiuCL18} and search~\cite{DBLP:conf/wsdm/WuLZWZM19, DBLP:conf/wsdm/KordanK18} tasks. The intuition is that the visual appearance of product images influences customer's decisions (e.g., a customer who loves red colors will likely buy red clothes)~\cite{DBLP:conf/wsdm/Grauman20}. For instance, He \textit{et al.}~\cite{DBLP:conf/aaai/HeM16} extended BPR-MF~\cite{DBLP:conf/uai/RendleFGS09} by integrating high-level features extracted from a pre-trained CNN, and Wang \textit{et al.}~\cite{DBLP:conf/icdm/KangFWM17} used image features to predict complementary fashion articles. Chu \textit{et al.}, and Wang \textit{et al.}~\cite{DBLP:conf/www/WangWTSRL17, DBLP:journals/www/ChuT17} demonstrated significantly improvements in POI-recommendations when considering food images features. Recently, Zhang and Caverlee~\cite{DBLP:conf/cikm/ZhangC19} proposed a novel VRS showing that dynamic visual features based on fashion blogger posts bring improvements in fashion recommendations.

However, recommender models have been demonstrated to be steadily under security risks.  The security of RSs relates to the study of different hand-engineered strategies to generate shilling profiles which lead to the alteration of collaborative recommendations~\cite{DBLP:conf/www/LamR04}, and their defense mechanisms (e.g., detection~\cite{bhaumik2006securing} and robustness~\cite{DBLP:journals/toit/OMahonyHKS04}). On the other hand, the application of AML in RSs~\cite{DBLP:conf/wsdm/DeldjooNM20} differs from previous works in the use of optimized perturbations, and their respective defenses, that lead to drastic performance reduction~\cite{DBLP:conf/sigir/0001HDC18, 8618394, DBLP:conf/wsdm/TangLSYMW20}. For example, He \textit{et al.}~\cite{DBLP:conf/sigir/0001HDC18} proposed an adversarial training procedure to make the model robust to such perturbations. Furthermore, Tang \textit{et al.}~\cite{8618394} applied this defense to make the proposed VRS (i.e., AMR) more robust to adversarial perturbations on image features. However, Di Noia \textit{et al.}~\cite{DSML20} noticed the partial protection of VBPR and AMR against targeted adversarial attacks on product images. Differently from these works, we empirically verified DNN robustification strategies are not always able to protect the recommender models against strong adversarial attacks (e.g., C \& W on all) when they are used to robustify the IFE.

\section{The Proposed Framework}\label{sec:proposed_framewok}
\begin{figure*}[!t]
    \centering
    \includegraphics[width=0.75\linewidth]{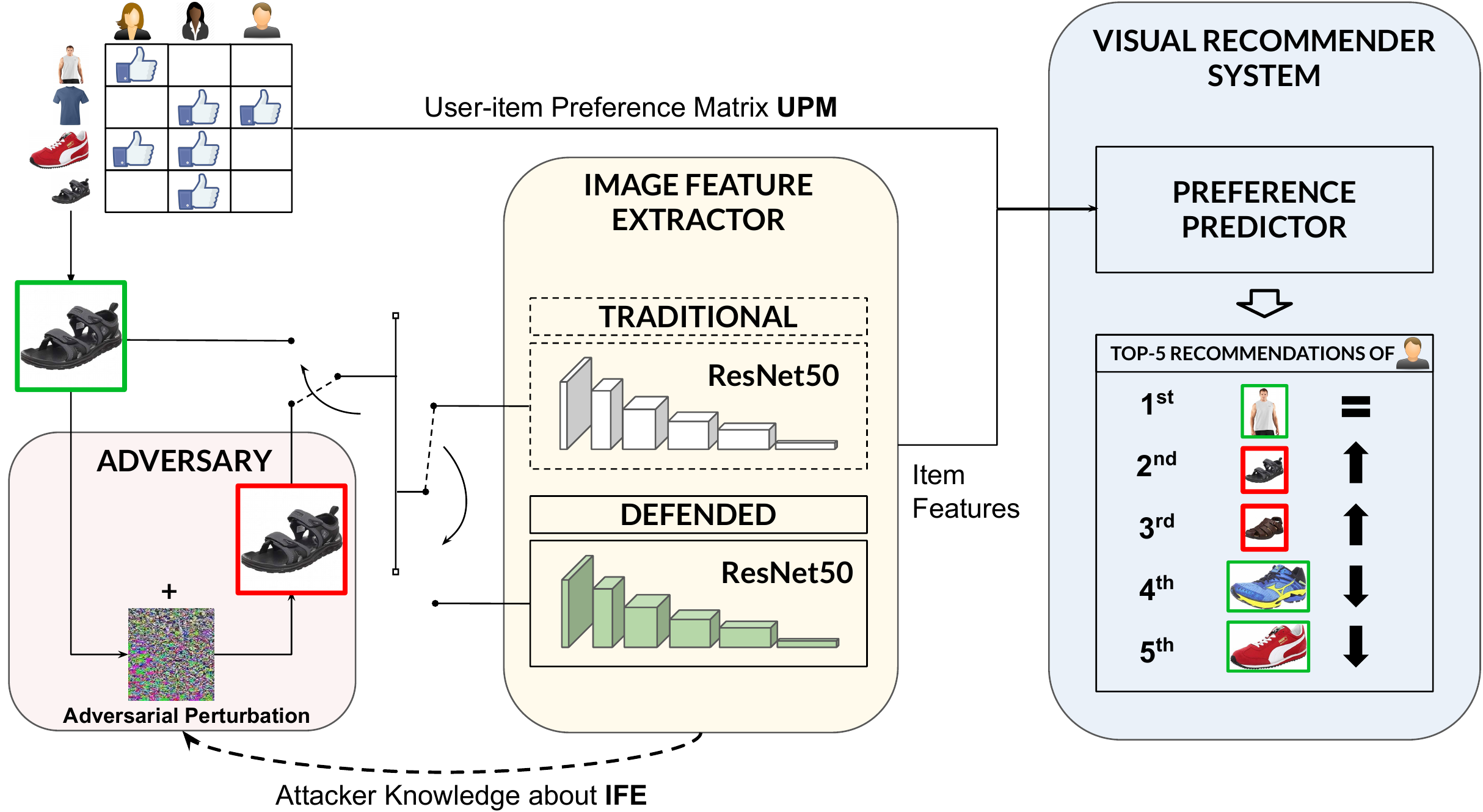}
    \caption{Overview of the VAR framework for the evaluation of adversarial attacks and defenses effects on a VRS. The adversary can perturb a product image, the Image Feature Extractor (IFE) extract the image visual features, and the Visual-based Recommender System (VRS) gets in input the user-item preference matrix (UPM) and the features to compute the top-$K$ lists.}
    \label{fig:var_architecture}
\end{figure*}

In this section, we describe the proposed Visual Adversarial Recommendation (\var) experimental framework.
First, we define some preliminary concepts. Then, we provide an overview on all \var components. Finally, we present the evaluation measures to quantify the effectiveness of the adversarial defenses under attacks.

\vspace{-0.25cm}

\subsection{Preliminaries}
We introduce some notions and notation to formalize \var. 

\textbf{Recommendation Task.} We define the set of users, items and $0/1$-valued preference feedback as $\mathcal{U}$, $\mathcal{I}$, and $\mathcal{S}$, where $|\mathcal{U}|$, $|\mathcal{I}|$, and $|\mathcal{S}|$ are the set sizes respectively. We reserve the use of $u$, $i$, and $s_{ui}$, to indicate a user in $\mathcal{U}$, an item in $\mathcal{I}$ and the feedback (e.g., a review) given by $u$ to $i$ saved in $\mathcal{S}$. Furthermore, we define the recommendation task as the action to suggest items that maximize, for each user, a utility function. We indicate with $\hat{s}_{ui}$ the predicted score learned from the RS upon historical preferences, represented as a user-item preference-feedback matrix (UPM).

\textbf{Deep Neural Network.} Given a set of data samples $(x_i, y_i)$, where $x_i$ is the $i$-th image and $y_i$ is the one-hot encoded representation of $x_i$'s image category, we define $F$ as a DNN classifier function trained on all $(x_i, y_i)$. Then, we set $F(x_i) = \hat{y}_i$ as the predicted probability vector of $x_i$ belonging to each of all the admissible classes, and we calculate its predicted class as the index of $\hat{y}_i$ with maximum probability value, and represent it as $F_c(x_i)$. Moreover, assuming an $L$-layers DNN classifier, we indicate with $F^{(l)}(x_i)$ the output of the $l$-th layer of $F$ given the input $x_i$.

\textbf{Adversarial Attack and Defense.} We define an \textit{Adversarial attack} as the problem of finding the best value for a perturbation $\delta_i$ such that:
\begin{align}
    \label{eq:adversarial_attack}
    \begin{aligned}
    \text{minimize } &\mathit{d}(x_i, x_i + \delta_i) \\
     F_c(x_i + \delta_i) &\neq F_c(x_i)\\  
    x_i + \delta_i &\in [0, 1]^n
    \end{aligned}
\end{align}
where $\mathit{d}(\cdot)$ is a \textit{distance metric} function (e.g., $L_0$, $L_2$ and $L_{\infty}$ norms). The above definition states that (i) the attacked image $x_i^{*} = x_i + \delta_i$ must be visually similar to $x_i$, (ii) the predicted class for $x_i^{*}$ must be different from the original one and (iii) $x_i^{*}$ must stay within its original value range (i.e., $[0, 1]$ for 8-bit RGB images re-scaled by a factor 255). When $F_c(x_i^{*})$ is required to be generically different from $F_c(x_i)$, we say the attack is \textit{untargeted}. On the contrary, when $F_c(x_i^{*})$ is specifically required to be equal to a target class $t$, we say the attack is \textit{targeted}.
Finally, we define a \textit{Defense} as the problem to find ways of limiting the impact of adversarial attacks against a DNN. For instance, a common solution consists of training a more robust version of the original model function ---we will refer to it as $\widetilde{F}$--- which attempts to classify attacked images correctly.

\subsection{Empirical Framework}\label{sec:framework}
After the definition of preliminaries, we discuss the \var components.  Figure~\ref{fig:var_architecture} shows an overview on the three main elements: the adversary (i.e., a malicious user), the image feature extractor (IFE), and the visual-based recommender system (VRS).  First, we describe the main characteristics of each mentioned component. Then, we introduce a novel set of metrics to evaluate the recommendation performance at varying of the adversary, the IFE, and the VRS. 

\textbf{Adversary.}
To align with the AML literature, we follow the attack ---and defense--- adversary threat model outlined in Carlini \textit{et al.}~\cite{DBLP:journals/corr/abs-1902-06705}. The \textit{adversary's goal} is to attack the IFE so that images of low-ranked categories of products are incorrectly classified as the category of high-ranked ones. That is, the former will likely be recommended more (on average) than before.
To this malicious purpose, the adversary is \textit{aware} of all recommendation lists (used to choose the source and target categories), and she has perfect knowledge of the IFE, i.e., its architecture, its trainable weights, and its output (white-box attack). Additionally, we suppose the adversary can perform $L_{\infty}$ (i.e., FGSM and PGD), and $L_{2}$ (i.e., Carlini \& Wagner) attacks (see section \ref{subsec:adv_attacks} for further details). Finally, in our motivating scenario, the adversary can \textit{replace the original images} on the physical servers of the e-commerce platform with the attacked one, which will be used by the VRS to produce the recommendation lists.

\textbf{Image Feature Extractor (IFE).}
The input sample $x_{i}$ represents the photo associated with the $i$-th product in the set of items $\mathcal{I}$, which may appear in the top-$K$ recommendation list shown to an e-commerce platform customer. Hence, the IFE is a DNN pre-trained classifier to extract high-level visual features from $x_i$. The actual extraction takes place at one of the last layers of the network, i.e., $F^{(e)}(x_i)$, where $e$ refers to the extraction layer. We define this layer output $F^{(e)}(x_i) = \varphi_{i}$ as a $\gamma$-dimensional vector that will be the input of the VRS. In the case of defense, the IFE model function is replaced by $\widetilde{F}$ since the defense strategy is applied to the pre-trained traditional model previously indicated as $F$. Note that the IFE is a key component in \var since it represents the connection between the adversary ---the author of the attack--- and the VRS.

\textbf{Visual-based Recommender System (VRS).}
In the \var framework, the VRS is the component aimed at addressing the recommendation task. The model accepts two inputs: (i) the historical UPM, and (ii) the set of features of item images extracted from the IFE component. Hence, it produces recommendation lists sorted by the preference prediction score evaluated for each user-item pair without previous interactions. Indeed, the VRS preference predictor takes advantage of the pure collaborative filtering source of data (i.e., the UPM) especially when integrated with the high-level multimedia features since they unveil the visual aspects that arouse customer's preferences~\cite{DBLP:conf/aaai/HeM16}.
In the \var motivating example, the VRS is the final victim of the malicious user. For this reason, we focus our analysis on its variation performance and propose novel measures to investigate how much it is influenced by different settings of both the adversary and the IFE.

\subsection{Evaluation}
To answer the research questions proposed in Section~\ref{sec:introduction}, we need to perform three levels of investigation on (i) the effectiveness of adversarial attacks in misusing the classification performance of the DNN used to implement the IFE, (ii) the variation of the accuracy--- and beyond-accuracy--- recommendation performance, and (iii) the evaluation of the consequences of attack and defense mechanisms on the recommendability of the attacked category of products.

In AML, several publications focused on quantifying adversarial attacks success in corrupting the classification performance of a target classifier (i.e., the attack Success Rate ($SR$)~\cite{DBLP:conf/sp/Carlini017}). Similarly, there is a vast literature about accuracy and beyond accuracy of RSs~\cite{DBLP:reference/sp/2015rsh} recommendation metrics. On the other hand, we have observed a lack of literature in evaluating adversarial attacks on RSs content data. As a matter of facts, Tang \textit{et al.}~\cite{8618394} evaluate the effects of untargeted attacks on classical system accuracy metrics, i.e., \textit{Hit Ratio} ($HR$) and \textit{normalized Discounted Cumulative Gain} ($nDCG$), while Di Noia \textit{et al.}~\cite{DSML20} propose a modified version of $HR$ to evaluate the fraction of adversarially perturbed items in the top-$K$ recommendations. To fill this evaluation metrics gap, we redefined the Category Hit Ratio (\chr)~\cite{DSML20} and formalized the normalized Category Discounted Cumulative Gain (\cndcg).  

\begin{definition}[\categoryhitratio]
Let $\mathcal{C}$ be the set of the classes extracted from the IFE, let $\mathcal{I}_{c} =  \{ i \in \mathcal{I}, c \in \mathcal{C} | F_c(x_{i})=c\}$ be the set of items whose images are classified by the IFE in the $c$-class (i.e., the category of low recommended items), we define categorical hit (chit) as:
\begin{equation}
        chit(u, k) = \begin{cases}
        1, & \text{if } \text{\textit{k-th item}} \in \mathcal{I}_{c}\\
        0, & \text{if } \text{\textit{k-th item}} \not\in \mathcal{I}_{c}
        \end{cases}
\end{equation}
where \textit{categorical hit} ($chit(u, k)$) is a $0$/$1$-valued function that is $1$ when the item in the $k$-th position of the top-$K$ recommendation list of the user $u$ is in the set of attacked items not-interacted by $u$.
Consequently, we define the \chr as follows:
    \begin{equation}
        \label{eq:chr}
        \centering
        CHR_u@K =  \frac{1}{K} \sum_{k = 1}^{K} chit(u, k)
    \end{equation}
\end{definition}

\noindent Since \textit{Category Hit Ratio} does not pay attention to the ranking of recommended items, we propose a novel rank-wise positional metric, named \textbf{Category normalized Discounted Cumulative Gain}, that assigns a gain to each considered ranking position.
By considering a relevance threshold $\tau$, we assume that each item $i \in \mathcal{I}_{c}$ has an ideal relevance value of:
\begin{equation}
idealrel(i) = 2^{(s_{max} - \tau + 1)} - 1
\end{equation}
where $s_{max}$ is the maximum possible score for items.
By considering a recommendation list provided to user $u$, we define the relevance ($rel(\cdot)$) of a suggested item $i$ as:
\begin{equation}
rel(k) = \begin{cases}
        2^{(s_{ui} - \tau + 1)} - 1, & \text{if } \text{\textit{k-th item}} \in \mathcal{I}_{c}\\
        0, & \text{if } \text{\textit{k-th item}} \not\in \mathcal{I}_{c}
        \end{cases}
\end{equation}
where $k$ is the position of the item $i$ in the recommendation list.
In Information Retrieval, the \textit{Discounted Cumulative Gain} ($DCG$) is a metric of ranking quality that measures the usefulness of a document based on its relevance and its position in the result list. 
Analogously, we define Category Discounted Cumulative Gain ($CDCG$) as:
\begin{equation}
	CDCG_{u}@K = \sum_{k=1}^{K} \frac{rel(k)}{\log_{2}(1+k)}
\end{equation}
Since recommendation results may vary in length depending on the user, it is not possible to compare performance among different users, so the cumulative gain at each position should be normalized across users.
In this respect, we define the \textit{Ideal Category Discounted Cumulative Gain} ($ICDCG@K$) as follows:
\begin{equation}
	ICDCG@K = \sum_{k=1}^{min(K, |\mathcal{I}_{c}|)} \frac{rel(k)}{\log_{2}(1+k)}
\end{equation}
In practical terms, $ICDCG@N$ indicates the score obtained by an ideal recommendation list that contains only relevant items.
\begin{definition}[normalized Category Discounted Cumulative Gain]
Let $\mathcal{C}$ be the set of the classes extracted from the IFE, let $\mathcal{I}_{c} =  \{ i \in \mathcal{I}, c \in \mathcal{C} | F_c(x_{i})=c\}$ be the set of items whose images are classified by the IFE in the $c$-class (i.e., the category of low recommended items).
Let $rel(k)$ be a function computing the relevance of the $k$-th item of the top-$K$ recommendation list, and $ICDCG@K$ be the $CDCG$ for an ideal recommendation list only composed of relevant items.
We define the \textit{normalized Category Discounted Cumulative Gain} ($nCDCG$), as:
\begin{equation}
	nCDCG_{u}@K = \frac{1}{ICDCG@K} \sum_{k=1}^{K} \frac{rel(k)}{\log_{2}(1+k)}
\end{equation}
\end{definition}
\noindent The \cndcg is ranged in a $[0, 1]$ interval, where values close to $1$ mean that the attacked items are recommended in higher positions (e.g., the attack is effective). In Information Retrieval, a logarithm with base $2$ is commonly adopted to ensure that all the recommendation list positions are discounted.

\section{Experimental Setup}\label{sec:experiment}


In this section, we first introduce the three real-world datasets, the adversarial attack strategies, the defense methods to make the IFE more robust, and the VRSs. Conclusively, we present the complete set of evaluation measures and a detailed presentation of the experimental choices to make the results reproducible.

\subsection{Datasets}

\textbf{Amazon Men \& Amazon Women}~\cite{DBLP:conf/www/HeM16} are two datasets about men's and women's products belonging to Amazon category "Clothing, Shoes and Jewelry". They come with both users' ratings and item images. Since we consider an implicit feedback setting, we transformed each user's rating into an implicit $0/1$-feedback.

\textbf{Tradesy}~\cite{DBLP:conf/aaai/HeM16} dataset contains implicit feedback (i.e., purchase histories, and desired products) extracted from the second-hand selling social platform of the same name. We followed the same pre-processing procedure seen for \amazonmen and \amazonwomen.

Moreover, to reduce the degrading effects of cold-users and items on recommendation performance, we applied different \textbf{\textit{k-core}} settings~\cite{DBLP:conf/www/HeM16}. In particular, we chose different \textbf{\textit{k}} values to explore various $\text{density}$ dataset characteristic settings. Table~\ref{table:dataset_statistics} shows the dataset statistics as a result of the pre-processing steps described above. The datasets are available on the code repository web page.
\begin{table}[!h]
    \centering
    \caption{Dataset statistics.}
    \scalebox{0.80}{
    \begin{tabular}{l l c c c c}
        \toprule
        \textbf{Dataset} & \textbf{\textit{k-cores}} &  \multicolumn{1}{c}{$|\mathcal{U}|$} & \multicolumn{1}{c}{$|\mathcal{I}|$} & \multicolumn{1}{c}{$|\mathcal{S}|$} & \multicolumn{1}{c}{$\text{density}$} \\
        \bottomrule
        \multicolumn{1}{l}{\amazonmen } & \textbf{5} & $24,379$ & $7,371$  & $ 89,020 $ & $0.000495$ \\ 
        \multicolumn{1}{l }{\amazonwomen } & \textbf{10} & $ 16,668 $ & $ 2,981$ & $ 54,473 $ & $0.001096$ \\ 
        \multicolumn{1}{l }{ \tradesy }  & \textbf{10} & $ 6,253 $ & $ 1,670 $ & $ 21,533 $ & $0.002062$ \\ 
        
        \bottomrule
    \end{tabular}
    }
    \label{table:dataset_statistics}
\end{table}

\subsection{Adversarial Attacks and Defenses}~\label{subsec:adv_attacks} 
In this section we present all the adversarial attack and defense techniques adopted in the experimental phase.

\subsubsection{Attacks}\label{subsubsection:attacks}
We explored three state-of-the-art adversarial attacks against images.

\textbf{Fast Gradient Sign Method (FGSM)}~\cite{DBLP:journals/corr/GoodfellowSS14}  is an $L_{\infty}$-norm optimized attack that produces an adversarial version of a given image in just one evaluation step. A perturbation budget $\epsilon$ is set to modify the strength ---and consequently, the visual perceptibility--- of the attack, i.e., higher $\epsilon$ values mean stronger attacks but also more evident visual artifacts.

\textbf{Projected Gradient Descent (PGD)}~\cite{DBLP:conf/iclr/MadryMSTV18} is an $L_{\infty}$-norm optimized attack that takes a uniform random noise as the initial perturbation, and \textit{iteratively} applies an FGSM attack with a continuously updated small perturbation $\alpha$ ---clipped within the $\epsilon$-ball--- until either it effectively reaches the network misclassification (i.e., $F_c(x_{i}+\alpha_{i}) = t$) or it completes the number of possible iterations (i.e., $10$ iterations in our evaluation setting).

\textbf{Carlini and Wagner attacks (C \& W)}~\cite{DBLP:conf/sp/Carlini017} are three attack strategies based on $L_0$, $L_2$ and $L_{\infty}$ norms that re-formulate the traditional adversarial attack problem (see \ref{eq:adversarial_attack}) by replacing the distance metric with a well-chosen \textit{objective function}. This integrates a parameter $\kappa$, i.e., the \textit{confidence} of the attacked image being classified as $t$, and an additional parameter $a$, i.e., the trade-off between optimizing the objective function and the classifier loss function. 

\subsubsection{Defenses}
We explored two defense strategies.

\textbf{Adversarial Training}~\cite{DBLP:journals/corr/GoodfellowSS14} consists of injecting adversarial samples into the training set to make the trained model robust to them. The major limitations of this idea are that it increases the computational time to complete the training phase, and it is deeply dependent on the type of attack strategy used to craft adversarial samples. For instance, Madry \textit{et al.}~\cite{DBLP:conf/iclr/MadryMSTV18} generates adversarial images with the PGD-method to make the trained model robust against both one-step and multi-steps attack strategies.

\textbf{Free Adversarial Training}~\cite{DBLP:conf/nips/ShafahiNG0DSDTG19} proposes a training procedure $3-30$ times faster than the classical Adversarial Training~\cite{DBLP:journals/corr/GoodfellowSS14, DBLP:conf/iclr/MadryMSTV18}. Differently from the previous one, this method updates both the model parameters and the adversarial perturbations doing a unique backward pass in which gradients are computed on the network loss. Moreover, to simulate a multi-step attack ---which would make the trained network more robust--- it keeps retraining on the same minibatch for $m$ times in a row. 

\subsection{Visual-based Recommender Models}~\label{subsec:recsys}
To evaluate \var approach, we have considered three VRSs.

\textbf{Factorization Machine (FM)}~\cite{DBLP:conf/icdm/Rendle10} is a recommender model proposed by Rendle~\cite{DBLP:conf/icdm/Rendle10} to estimate the user-item preference score with a factorization technique. For a fair comparison with VBPR and AMR, we used BPR~\cite{DBLP:conf/uai/RendleFGS09} loss function to optimize the personalized ranking. In this respect, we adopted LightFM~\cite{DBLP:conf/recsys/Kula15} implementation integrating the UPM with the extracted continuous features. It is worth noticing that, differently from the recommenders we are going to present later, this model is not specifically designed for visual-recommendation tasks.

\textbf{Visual Bayesian Personalized Ranking (VBPR)}~\cite{DBLP:conf/aaai/HeM16} is a typical matrix factorization CF model to learn user-item latent representation by optimizing a BPR rank-wise loss function. Given a user $u$ and a not-interacted item $i$, the preference prediction score is $ \hat{s}_{ui}= p_{u}^{T} q_{i} + \theta_{u}^{T}(\mathbf{E} \varphi_{i}) + \beta_{ui}$, where $p_{u} \in \mathbb{P}^{|\mathcal{U}| \times h}$ and $q_{i} \in \mathbb{Q}^{|\mathcal{I}| \times h}$ are latent vectors of user $u$ and item $i$ respectively, $h$ is the latent space dimension ($h << |\mathcal{U}|, |\mathcal{I}| $), and $\theta_{u}$ is a $\upsilon$-dimensional vector to capture the visual interaction between $u$ and the projection of $\varphi_i$ into a low-dimensional space through a $(\upsilon \times \gamma)$-kernel matrix $\mathbf{E}$. Furthermore, $\beta_{ui}$ includes the sum of the overall offset, and the user, item and global visual bias.

\textbf{Adversarial Multimedia Recommendation (AMR)}~\cite{8618394} is an extension of VBPR that integrates the adversarial training procedure proposed by He \textit{et al.}~\cite{DBLP:conf/sigir/0001HDC18} named \textit{adversarial regularization} to build a model that is increasingly robust to FGSM-based perturbations against image features. The score prediction function is the same as VBPR since the differences are included in the training procedure.

\subsection{Evaluation Metrics}
In addition to \chr and \cndcg ---proposed in Section~\ref{sec:proposed_framewok}--- 
, we studied both the variation of overall recommendation performance and the consequences of adversarial images on th IFE.

\textbf{Adversarial attacks, and defenses, performance} are evaluated through the attack Success Rate ($SR$), and the Feature Loss ($FL$), i.e., the mean squared error between the extracted image features before and after the attack.

\textbf{Recommendation performance} is evaluated with $Recall@K$, that considers the fraction of recommended products in the top-$K$ recommendation that hit test items, and $nDCG@K$, that increasingly discounts the hits by the $\log_2$ of the item positions in the list. Moreover, we investigate three \textit{beyond-accuracy} measures: the item coverage ($ICov@K$), the Gini index ($Gini@K$), and the expected free discovery ($EFD@K$).
Please note that \chr, \cndcg, $Recall@K$, $nDCG@K$, and $EFD@K$ are computed on a per-user basis, and then averaged across all users.


\subsection{Reproducibility}
To let other researchers to continue our study, and further integrate the framework, in this section, we provide reproducibility details.

\textbf{Adversarial attacks} were implemented with the Python library CleverHans~\cite{papernot2018cleverhans}. For both FGSM and PGD, we adopted $\epsilon = \{4, 8\}$ re-scaled by 255. Then, for PGD's $\alpha$ parameter, we set the multi-step size as $\epsilon/6$ and the number of iterations to 10.
As for the C \& W attack, we ran a 5-steps binary search to calculate $a$, starting from an initial value of $10^{-2}$ and set $\kappa$ to 0. Furthermore, we set the maximum number of iteration to 1000 and adopted Adam optimizer with a learning rate of $5 \times 10^{-3}$ as suggested in C \& W~\cite{DBLP:conf/sp/Carlini017}. 
Note that, to reproduce a real attack scenario, we saved the adversarial images in \texttt{tiff} format (i.e., a lossless compression), as lossy compression (e.g., JPEG) may affect the effectiveness of attacks~\cite{DBLP:conf/iclr/GuoRCM18}.

\textbf{Feature extraction and Defenses.} We used ResNet50~\cite{DBLP:conf/cvpr/HeZRS16} to extract high-level image features. From the PyTorch implementation, we set \texttt{AdaptiveAvgPool2d} as extraction layer, whose output is a $2048$-dimensional vector. In the non-defended scenario, we adopted ResNet50 pre-trained on \imagenet with traditional training. Conversely, when applying defense techniques, we adopted ResNet50 pre-trained on \imagenet with Adversarial Training and Free Adversarial Training. For the former, we used a model trained with $\epsilon=4$. For the latter, we used a model trained with $\epsilon=4$ and $m = 4$. Both models are available in the published repository.

\textbf{Recommenders.} We realized the FM model using the LightFM library~\cite{DBLP:conf/recsys/Kula15}. We trained the model for 100 epochs and left all the parameters with the default values in the library. Both VBPR and AMR were implemented in \textit{TensorFlow}. We trained the models following the training settings adopted in Tang \textit{et al.}~\cite{8618394}. We did not apply the hyper-parameters search, and used the parameters suggested in the referenced works, since the goal of our evaluation is to investigate the protection abilities of defense mechanisms against attacks by fixing a VRS.

\textbf{Experimental Scenario.} \textit{We trained each recommender on clean images, we selected the origin-target categories such that target ones was about four times more recommended of the origin one, and trained a novel model using the perturbed images.} Table~\ref{table:category} shows the selected categories.  For each dataset, we used the \textit{leave-one-out} training-test protocol putting in the test set the last time-aware user's interaction.

\begin{table}[!t]
    \centering
    \caption{Origin-target category classes selected for the \var experimental evaluation. \chratcinq is averaged across the 9 combinations of recommenders and defenses without attacks.}
    \scalebox{0.80}{
    \begin{tabular}{l l c l c}
        \toprule
        \textbf{Dataset} & \textbf{\textit{Origin}} &  \multicolumn{1}{c}{\chratcinq} & \textbf{\textit{Target}} &  \multicolumn{1}{c}{\chratcinq} \\
        \bottomrule
        \multicolumn{1}{l}{\amazonmen }     & Sandal & 1.0310 & Running Shoe & 4.7852\\ 
        \multicolumn{1}{l }{\amazonwomen }  & Jersey, T-shirt & 1.2573 & Brassiere, Bandeau & 4.2672\\ 
        \multicolumn{1}{l }{\tradesy }      & Suit & 0.8951 & Trench Coat & 3.6955\\ 
        
        \bottomrule
    \end{tabular}}
    
    \label{table:category}
\end{table}

\section{Discussion of the Results}\label{sec:discussion}
\begin{table}[!t]
    \caption{Average values of Success Rate ($SR$) and Feature Loss ($FL$) for each combination. $SR$ values are multiplied by $10^{-3}$.}
\scalebox{0.75}{
\begin{tabular}{clcr|cr|cr}
\bottomrule
\multirow{4}{*}{\textbf{Dataset}} & \multirow{4}{*}{\begin{tabular}[r]{@{}c@{}}\textbf{Adversarial}\\\textbf{Attack}\end{tabular}} & \multicolumn{ 6}{c}{\textbf{Image Feature Extractor}} \\ \cmidrule(lr){3-8}
\multicolumn{ 1}{c}{} & \multicolumn{ 1}{c}{} & \multicolumn{ 2}{c}{\textbf{Traditional}} & \multicolumn{ 2}{c}{\textbf{Adv. Train.}} & \multicolumn{ 2}{c}{\textbf{Free Adv. Train.}} \\
\cmidrule(lr){3-4} \cmidrule(lr){5-6} \cmidrule(lr){7-8}
\multicolumn{ 1}{c}{} & \multicolumn{ 1}{c}{} &
\multicolumn{ 1}{c}{$SR$} &
\multicolumn{ 1}{c}{$FL$}  &
\multicolumn{ 1}{c}{$SR$} &
\multicolumn{1}{c}{$FL$}  &
\multicolumn{ 1}{c}{$SR$} &
\multicolumn{1}{c}{$FL$}\\ \bottomrule

\multirow{5}{*}{\begin{tabular}[l]{@{}c@{}}\texttt{Amazon}\\\texttt{Men}\end{tabular}} 
& FGSM ($\epsilon = 4$) & 65\% & \textbf{14.0948}    & 18\% & \textbf{0.0330} & 15\% & \textbf{0.0278}  \\

\multicolumn{ 1}{c}{}   & FGSM ($\epsilon = 8$) & 87\% & 36.3190  & 24\% & 0.2658 & 20\% & 0.2320 \\ 

\multicolumn{ 1}{c}{}   & PGD ($\epsilon = 4$)  & 97\% & 36.8843    & 18\% & 0.0334  & 15\% & 0.0283  \\ 

\multicolumn{ 1}{c}{}   & PGD ($\epsilon = 8$)  & \textbf{100\%} & 134.9854 & 24\% & 0.2801  & 21\% & 0.2371 \\

\multicolumn{ 1}{c}{}   & C \& W & 89\% & 20.5172  & \textbf{48\%} & 2.8022  & \textbf{42\%} & 1.9080 \\ \hline

\multirow{5}{*}{\begin{tabular}[l]{@{}c@{}}\texttt{Amazon}\\\texttt{Women}\end{tabular}} 
                      & FGSM ($\epsilon = 4$) & 18\% & \textbf{9.6677} & 0\% & \textbf{0.0113} & 0\% &  \textbf{0.0094} \\

\multicolumn{ 1}{c}{} & FGSM ($\epsilon = 8$) & 28\% & 22.0499 & 3\% & 0.0851 & 0\% & 0.0671 \\ 

\multicolumn{ 1}{c}{} & PGD ($\epsilon = 4$) & 85\% & 27.6645 & 0\% & 0.0119 & 0\% & 0.0102 \\

\multicolumn{ 1}{c}{} & PGD ($\epsilon = 8$) & \textbf{100\%} & 130.3309 & 4\% & 0.0974 & 0\% & 0.0735 \\ 

\multicolumn{ 1}{c}{} & C \& W & 89\% & 21.2380 & \textbf{6\%} & 0.1770  & \textbf{6\%} & 0.3376  \\ \hline

\multirow{5}{*}{\begin{tabular}[l]{@{}c@{}}\texttt{Tradesy}\end{tabular}} 

                      & FGSM ($\epsilon = 4$) & 83\% & \textbf{21.4011}  & 43\% & \textbf{0.0308}  & 30\% & 0.0274  \\

\multicolumn{ 1}{c}{} & FGSM ($\epsilon = 8$) & 93\% & 46.2579 & 47\% & 0.2376 & 47\% & 0.2130 \\

\multicolumn{ 1}{c}{} & PGD ($\epsilon = 4$) & \textbf{100\%} & 53.4589  & 43\% & 0.0311 & 30\% & \textbf{0.0273} \\

\multicolumn{ 1}{c}{} & PGD ($\epsilon = 8$) & \textbf{100\%} & 175.7102  & 47\% & 0.2478 & 47\% & 0.2078 \\

\multicolumn{ 1}{c}{} & C \& W & \textbf{100\%} & 25.9374 & \textbf{80\%} & 2.1185  & \textbf{63\%} & 1.9739 \\ \bottomrule
\end{tabular}
\label{table:image_metrics}
}
\end{table}

\begin{table*}[!th]
\caption{Results of the \var framework. Bold values are the highest values for each <dataset, VRS, defense> combination.}

\resizebox{0.95\textwidth}{!}{

\begin{tabular}{ l l l  c  c  c  c| c  c  c  c | c  c  c  c }

\toprule

\multirow{4}{*}{\textbf{Dataset}}            &  \multirow{4}{*}{\textbf{VRS}}            &  \multirow{4}{*}{\begin{tabular}[r]{@{}c@{}}\textbf{Adversarial}\\\textbf{Attack}\end{tabular}}            & \multicolumn{12}{c}{\textbf{Image Feature Extractor}}

\\ 

\cmidrule(lr){4-15} 

\multicolumn{ 1}{c}{}            & \multicolumn{ 1}{c}{}            & \multicolumn{ 1}{c}{}            & \multicolumn{ 4}{c}{\textbf{Traditional}}            & \multicolumn{ 4}{c}{\textbf{Adversarial Training}}            & \multicolumn{ 4}{c}{\textbf{Free Adversarial Training}} \\ 

\cmidrule(lr){4-7}\cmidrule(lr){8-11}\cmidrule(lr){12-15} 

\multicolumn{ 1}{l}{}& \multicolumn{ 1}{l}{}& \multicolumn{ 1}{l}{}&
\multicolumn{1}{c}{\footnotesize{\chratventi}}            &  \multicolumn{1}{c}{\footnotesize{\chratcinq}}            &    \multicolumn{1}{c}{\footnotesize{\cndcgatventi}}            & \multicolumn{1}{c}{\footnotesize{\cndcgatcinq}}            & \multicolumn{1}{c}{\footnotesize{\chratventi}}            &   \multicolumn{1}{c}{\footnotesize{\chratcinq}}            & \multicolumn{1}{c}{\footnotesize{\cndcgatventi}}            &\multicolumn{1}{c}{\footnotesize{\cndcgatcinq}}            &  \multicolumn{1}{c}{\footnotesize{\chratventi}}            & \multicolumn{1}{c}{\footnotesize{\chratcinq}}            &   \multicolumn{1}{c}{\footnotesize{\cndcgatventi}}            & \multicolumn{1}{c}{\footnotesize{\cndcgatcinq}} \\ 

\bottomrule
\multirow{18}{*}{\begin{tabular}[l]{@{}c@{}}\texttt{Amazon}\\\texttt{Men}\end{tabular}}  
          & \multirow{6}{*}{FM}               & Original                             & 0.4857            & 1.1959            & 0.0246            & 0.0245            & \textbf{0.4003}   & \textbf{0.9793}   & \textbf{0.0204}   & \textbf{0.0201}   & \textbf{0.3984}   & \textbf{0.9791}   & \textbf{0.0202}   & \textbf{0.0201}\\ 
\multicolumn{ 1}{l}{}& \multicolumn{ 1}{c}{}  & FGSM ($\epsilon=4$)                  & \textbf{0.5193}   & 1.2670            & \textbf{0.0266}   & 0.0262            & 0.3811            & 0.9315            & 0.0198            & 0.0192            & 0.3750            & 0.9324            & 0.0194            & 0.0192 \\ 
\multicolumn{ 1}{l}{}& \multicolumn{ 1}{l}{}  & FGSM ($\epsilon=8$)                  & 0.5092            & 1.2693            & 0.0263            & 0.0261            & 0.3715            & 0.9299            & 0.0193            & 0.0191            & 0.3837            & 0.9281            & 0.0195            & 0.0189 \\ 
\multicolumn{ 1}{l}{}& \multicolumn{ 1}{l}{}  & PGD ($\epsilon=4$)                   & 0.5147            & \textbf{1.2692}   & \textbf{0.0266}   & \textbf{0.0263}   & 0.3729            & 0.9187            & 0.0191            & 0.0188            & 0.3735            & 0.9112            & 0.0193            & 0.0188 \\ 
\multicolumn{ 1}{l}{}& \multicolumn{ 1}{l}{}  & PGD ($\epsilon=8$)                   & 0.5024            & 1.2672            & 0.0256            & 0.0256            & 0.3765            & 0.9215            & 0.0194            & 0.0190            & 0.3825            & 0.9367            & 0.0194            & 0.0191 \\ 
\multicolumn{ 1}{l}{}& \multicolumn{ 1}{l}{}  & C \& W                               & 0.5155            & 1.2573            & 0.0263            & 0.0261            & 0.3765            & 0.9338            & 0.0194            & 0.0191            & 0.3798            & 0.9428            & 0.0194            & 0.0192 \\ \cline{ 2- 15}

\multicolumn{ 1}{l}{}& \multirow{6}{*}{VBPR} & Original                              & 0.6352            & 1.4931            & 0.0288            & 0.0291            & 0.3028            & 0.8130            & 0.0141            & 0.0155            & 0.3702            & 1.1547            & 0.0159            & 0.0207 \\ 
\multicolumn{ 1}{l}{}& \multicolumn{ 1}{l}{}  & FGSM ($\epsilon=4$)                  & 0.5665            & 1.2607            & 0.0299            & 0.0269            & 0.6029            & 1.4496            & 0.0316            & 0.0306            & 0.5688            & 1.4924            & 0.0283            & 0.0296 \\ 
\multicolumn{ 1}{l}{}& \multicolumn{ 1}{l}{}  & FGSM ($\epsilon=8$)                  & 0.6052            & 1.3498            & 0.0342            & 0.0300            & 0.5879            & 1.4333            & 0.0316            & 0.0302            & 0.5596            & 1.4433            & 0.0277            & 0.0290 \\ 
\multicolumn{ 1}{l}{}& \multicolumn{ 1}{l}{}  & PGD ($\epsilon=4$)                   & 1.0936            & 2.6175            & 0.0538            & 0.0539            & 0.6211            & \textbf{1.4763}   & 0.0324            & 0.0309            & 0.5778            & 1.5003            & 0.0286            & 0.0301 \\ 
\multicolumn{ 1}{l}{}& \multicolumn{ 1}{l}{}  & PGD ($\epsilon=8$)                   & \textbf{1.5736}   & \textbf{3.7285}   & \textbf{0.0781}   & \textbf{0.0780}   & 0.6247            & 1.4565            & 0.0335            & \textbf{0.0312}   & 0.6141            & \textbf{1.5768}   & 0.0310            & 0.0320 \\ 
\multicolumn{ 1}{l}{}& \multicolumn{ 1}{l}{}  & C \& W                               & 0.5972            & 1.4003            & 0.0290            & 0.0285            & \textbf{0.6652}   & 1.4487            & \textbf{0.0336}   & 0.0305            & \textbf{0.6444}   & 1.5641            & \textbf{0.0348}   & \textbf{0.0334}\\ \cline{ 2- 15}

\multicolumn{ 1}{l}{}& \multirow{6}{*}{AMR}  & Original                              & 0.3876            & 0.8587            & 0.0196            & 0.0178            & 0.4924            & 1.1802            & 0.0228            & 0.0230            & 0.1070            & 0.6255            & 0.0038            & 0.0100 \\ 
\multicolumn{ 1}{l}{}& \multicolumn{ 1}{c}{}  & FGSM ($\epsilon=4$)                  & 0.3295            & 0.8282            & 0.0150            & 0.0159            & 0.4332            & 1.1736            & 0.0235            & 0.0242            & 0.4103            & \textbf{1.1595}   & 0.0187            & 0.0217 \\ 
\multicolumn{ 1}{l}{}& \multicolumn{ 1}{l}{}  & FGSM ($\epsilon=8$)                  & 0.3053            & 0.8668            & 0.0135            & 0.0160            & 0.4318            & \textbf{1.1827}   & 0.0238            & 0.0246            & 0.4007            & 1.1250            & 0.0188            & 0.0214 \\ 
\multicolumn{ 1}{l}{}& \multicolumn{ 1}{l}{}  & PGD ($\epsilon=4$)                   & 0.8064            & 1.9749            & 0.0418            & 0.0413            & \textbf{0.4435}   & 1.1756            & \textbf{0.0242}   & \textbf{0.0247}   & 0.4173            & 1.1657            & 0.0193            & 0.0220 \\ 
\multicolumn{ 1}{l}{}& \multicolumn{ 1}{l}{}  & PGD ($\epsilon=8$)                   & \textbf{2.1264}   & \textbf{5.2984}   & \textbf{0.1179}   & \textbf{0.1141}   & 0.4323            & 1.1447            & 0.0237            & 0.0241            & 0.3942            & 1.1386            & 0.0181            & 0.0213 \\ 
\multicolumn{ 1}{l}{}& \multicolumn{ 1}{l}{}  & C \& W                               & 0.3610            & 0.8171            & 0.0170            & 0.0163            & 0.4293            & 1.1227            & 0.0230            & 0.0233            & \textbf{0.4378}   & 1.1623            & \textbf{0.0202}   & \textbf{0.0224} \\ \hline

\multirow{18}{*}{\begin{tabular}[l]{@{}c@{}}\texttt{Amazon}\\\texttt{Women}\end{tabular}}              
& \multirow{6}{*}{FM}                         & Original                             & 0.6771            & 1.6409            & 0.0335            & 0.0333            & 0.4622            & 1.1589            & 0.0236            & 0.0237            & \textbf{0.3186}   & \textbf{0.7741}   & \textbf{0.0158}   & \textbf{0.0155}\\ 
\multicolumn{ 1}{l}{}& \multicolumn{ 1}{c}{}  & FGSM ($\epsilon=4$)                  & 0.6816            & 1.6805            & 0.0354            & 0.0351            & 0.4708            & 1.1550            & 0.0243            & 0.0239            & 0.2985            & 0.7369            & 0.0145            & 0.0144 \\ 
\multicolumn{ 1}{l}{}& \multicolumn{ 1}{l}{}  & FGSM ($\epsilon=8$)                  & 0.6880            & 1.6876            & 0.0355            & 0.0351            & 0.4730            & 1.1593            & 0.0242            & 0.0239            & 0.2985            & 0.7319            & 0.0143            & 0.0142 \\ 
\multicolumn{ 1}{l}{}& \multicolumn{ 1}{l}{}  & PGD ($\epsilon=4$)                   & \textbf{0.6900}   & \textbf{1.6920}   & \textbf{0.0356}   & \textbf{0.0352}   & 0.4737            & 1.1503            & 0.0244            & 0.0238            & 0.3057            & 0.7451            & 0.0155            & 0.0152 \\ 
\multicolumn{ 1}{l}{}& \multicolumn{ 1}{l}{}  & PGD ($\epsilon=8$)                   & 0.6727            & 1.6561            & 0.0348            & 0.0344            & \textbf{0.4919}   & \textbf{1.1811}   & \textbf{0.0254}   & \textbf{0.0246}   & 0.2988            & 0.7374            & 0.0151            & 0.0150 \\ 
\multicolumn{ 1}{l}{}& \multicolumn{ 1}{l}{}  & C \& W                               & 0.6746            & 1.6461            & 0.0329            & 0.0329            & 0.4655            & 1.1409            & 0.0240            & 0.0237            & 0.2844            & 0.7062            & 0.0144            & 0.0144 \\ \cline{ 2- 15}

\multicolumn{ 1}{l}{}& \multirow{6}{*}{VBPR} & Original                              & 0.4377            & 1.2812            & 0.0199            & 0.0237            & 0.5108            & 1.2390            & 0.0251            & 0.0251            & 0.3417            & 0.9570            & 0.0161            & 0.0184 \\ 
\multicolumn{ 1}{l}{}& \multicolumn{ 1}{l}{}  & FGSM ($\epsilon=4$)                  & 0.3860            & 1.0793            & 0.0174            & 0.0198            & 0.6032            & 1.3813            & 0.0310            & 0.0292            & 0.6088            & 1.1151            & 0.0303            & 0.0246 \\ 
\multicolumn{ 1}{l}{}& \multicolumn{ 1}{l}{}  & FGSM ($\epsilon=8$)                  & 0.4057            & 1.2445            & 0.0179            & 0.0228            & 0.6186            & 1.4160            & 0.0319            & 0.0301            & \textbf{0.6313}   & 1.1662            & \textbf{0.0332}   & \textbf{0.0263} \\ 
\multicolumn{ 1}{l}{}& \multicolumn{ 1}{l}{}  & PGD ($\epsilon=4$)                   & 0.9142            & 2.3673            & 0.0459            & 0.0483            & 0.6309            & 1.4456            & 0.0315            & 0.0296            & 0.6263            & 1.1165            & 0.0330            & 0.0257 \\ 
\multicolumn{ 1}{l}{}& \multicolumn{ 1}{l}{}  & PGD ($\epsilon=8$)                   & \textbf{1.4462}   & \textbf{3.4759}   & \textbf{0.0748}   & \textbf{0.0741}   & \textbf{0.6413}   & \textbf{1.4674}   & \textbf{0.0336}   & \textbf{0.0314}   & 0.6139            & \textbf{1.1194}   & \textbf{0.0322}   & 0.0254 \\ 
\multicolumn{ 1}{l}{}& \multicolumn{ 1}{l}{}  & C \& W                               & 0.4147            & 1.2121            & 0.0173            & 0.0214            & 0.6280            & 1.4277            & 0.0326            & 0.0303            & 0.5729            & 1.1019            & 0.0302            & 0.0247 \\ \cline{ 2- 15}

\multicolumn{ 1}{l}{}& \multirow{6}{*}{AMR}  & Original                              & 0.9449            & 2.0206            & 0.0462            & 0.0422            & \textbf{0.8342}   & 1.4602            & \textbf{0.0433}   & 0.0332            & \textbf{0.5063}   & 0.7841            & \textbf{0.0303}   & 0.0207 \\ 
\multicolumn{ 1}{l}{}& \multicolumn{ 1}{c}{}  & FGSM ($\epsilon=4$)                  & \textbf{1.3173}   & \textbf{2.4648}   & 0.0862            & \textbf{0.0649}   & 0.7135            & 1.7675            & 0.0334            & 0.0351            & 0.4565            & 1.0392            & 0.0230            & 0.0217 \\ 
\multicolumn{ 1}{l}{}& \multicolumn{ 1}{l}{}  & FGSM ($\epsilon=8$)                  & 1.2814            & 2.2121            & \textbf{0.0876}   & 0.0620            & 0.7137            & 1.7595            & 0.0341            & 0.0356            & 0.4429            & 1.0408            & 0.0221            & 0.0213 \\ 
\multicolumn{ 1}{l}{}& \multicolumn{ 1}{l}{}  & PGD ($\epsilon=4$)                   & 1.1958            & 2.0161            & 0.0713            & 0.0517            & 0.6473            & 1.7284            & 0.0307            & 0.0342            & 0.4900            & \textbf{1.0750}   & 0.0240            & \textbf{0.0219} \\ 
\multicolumn{ 1}{l}{}& \multicolumn{ 1}{l}{}  & PGD ($\epsilon=8$)                   & 1.2377            & 2.6192            & 0.0679            & 0.0593            & 0.6770            & 1.7451            & 0.0322            & 0.0346            & 0.4445            & 1.0364            & 0.0221            & 0.0213 \\ 
\multicolumn{ 1}{l}{}& \multicolumn{ 1}{l}{}  & C \& W                               & 1.3012            & 2.2742            & 0.0746            & 0.0558            & 0.7159            & \textbf{1.7976}   & 0.0336            & \textbf{0.0357}   & 0.4977            & 1.0714            & 0.0243            & \textbf{0.0219} \\ \hline

\multirow{18}{*}{\tradesy}            & \multirow{6}{*}{FM}  
                                              & Original                             & 0.3371            & 0.8935            & 0.0145            & 0.0217            & 0.3579            & 0.8852            & 0.0160            & 0.0219            & 0.4649            & 1.1398            & 0.0212            & 0.0246 \\ 
\multicolumn{ 1}{l}{}& \multicolumn{ 1}{c}{}  & FGSM ($\epsilon=4$)                  & 0.3617            & 0.9098            & 0.0161            & 0.0228            & 0.3744            & 0.9151            & 0.0168            & 0.0227            & 0.5118            & 1.2292            & 0.0236            & 0.0261 \\ 
\multicolumn{ 1}{l}{}& \multicolumn{ 1}{l}{}  & FGSM ($\epsilon=8$)                  & 0.3696            & 0.9232            & 0.0164            & 0.0229            & 0.3822            & 0.9141            & 0.0166            & 0.0218            & \textbf{0.5119}   & 1.2271            & 0.0233            & 0.0257 \\ 
\multicolumn{ 1}{l}{}& \multicolumn{ 1}{l}{}  & PGD ($\epsilon=4$)                   & 0.3603            & 0.9095            & 0.0158            & 0.0226            & 0.3598            & 0.9024            & 0.0158            & 0.0216            & 0.5081            & \textbf{1.2300}   & 0.0236            & 0.0256 \\ 
\multicolumn{ 1}{l}{}& \multicolumn{ 1}{l}{}  & PGD ($\epsilon=8$)                   & \textbf{0.4028}   & \textbf{0.9811}   & \textbf{0.0181}   & \textbf{0.0250}   & 0.3741            & 0.9018            & 0.0165            & 0.0221            & 0.5092            & 1.2295            & 0.0234            & 0.0259 \\ 
\multicolumn{ 1}{l}{}& \multicolumn{ 1}{l}{}  & C \& W                               & 0.3750            & 0.9242            & 0.0167            & 0.0234            & \textbf{0.3913}   & \textbf{0.9356}   & \textbf{0.0179}   & \textbf{0.0233}   & 0.5116            & 1.2260            & \textbf{0.0240}   & \textbf{0.0269} \\ \cline{ 2- 15}

\multicolumn{ 1}{l}{}& \multirow{6}{*}{VBPR} & Original                              & 0.4108            & 1.0721            & 0.0186            & 0.0271            & 0.2973            & 0.7526            & 0.0122            & 0.0173            & 0.3179            & 0.9602            & 0.0129            & 0.0180 \\ 
\multicolumn{ 1}{l}{}& \multicolumn{ 1}{l}{}  & FGSM ($\epsilon=4$)                  & 0.5202            & 1.2273            & 0.0260            & 0.0333            & \textbf{0.5055}   & 0.9778   & \textbf{0.0242}   & \textbf{0.0267}   & \textbf{0.5644}   & 1.1905            & \textbf{0.0261}   & \textbf{0.0267} \\ 
\multicolumn{ 1}{l}{}& \multicolumn{ 1}{l}{}  & FGSM ($\epsilon=8$)                  & 0.7251            & 1.4667            & 0.0408            & 0.0458            & 0.4807            & 0.9567            & 0.0224            & 0.0252            & 0.4868            & 1.0908            & 0.0210            & 0.0230 \\ 
\multicolumn{ 1}{l}{}& \multicolumn{ 1}{l}{}  & PGD ($\epsilon=4$)                   & 1.2552            & 2.2920            & 0.0649            & 0.0699            & 0.4431            & 0.8885            & 0.0199            & 0.0226            & 0.5217            & 1.1586            & 0.0235            & 0.0243 \\ 
\multicolumn{ 1}{l}{}& \multicolumn{ 1}{l}{}  & PGD ($\epsilon=8$)                   & \textbf{1.6982}   & \textbf{2.9039}   & \textbf{0.0913}   & \textbf{0.0971}   & 0.4726            & 0.9346            & 0.0211            & 0.0243            & 0.5317            & 1.1580            & 0.0232            & 0.0245 \\ 
\multicolumn{ 1}{l}{}& \multicolumn{ 1}{l}{}  & C \& W                               & 0.4523            & 1.0561            & 0.0221            & 0.0286            & 0.4766            & \textbf{0.9920}            & 0.0223            & 0.0257            & 0.5474            & \textbf{1.2309}   & 0.0242            & 0.0252 \\ \cline{ 2- 15}

\multicolumn{ 1}{l}{}& \multirow{6}{*}{AMR}  & Original                              & 0.3653            & 1.0573            & 0.0154            & 0.0252            & 0.1626            & 0.5346            & 0.0059            & 0.0115            & 0.2189            & 0.7609            & 0.0079            & 0.0135 \\ 
\multicolumn{ 1}{l}{}& \multicolumn{ 1}{c}{}  & FGSM ($\epsilon=4$)                  & 0.4759            & 1.2386            & 0.0215            & 0.0325            & \textbf{0.3587}   & 0.8059            & 0.0162            & 0.0202            & \textbf{0.4041}   & \textbf{0.9674}   & \textbf{0.0180}   & \textbf{0.0202} \\ 
\multicolumn{ 1}{l}{}& \multicolumn{ 1}{l}{}  & FGSM ($\epsilon=8$)                  & 0.5896            & 1.3448            & 0.0302            & 0.0376            & 0.3512            & 0.7887            & 0.0165            & 0.0200            & 0.3745            & 0.9132            & 0.0159            & 0.0184 \\ 
\multicolumn{ 1}{l}{}& \multicolumn{ 1}{l}{}  & PGD ($\epsilon=4$)                   & 1.0393            & 1.9715            & 0.0523            & 0.0593            & 0.3438            & 0.7686            & 0.0160            & 0.0193            & 0.3635            & 0.8986            & 0.0160            & 0.0184 \\ 
\multicolumn{ 1}{l}{}& \multicolumn{ 1}{l}{}  & PGD ($\epsilon=8$)                   & \textbf{1.6016}   & \textbf{2.6768}   & \textbf{0.0781}   & \textbf{0.0852}   & 0.3494            & 0.7726            & 0.0160            & 0.0197            & 0.3864            & 0.9416            & 0.0171            & 0.0193 \\ 
\multicolumn{ 1}{l}{}& \multicolumn{ 1}{l}{}  & C \& W                               & 0.4302            & 1.1161            & 0.0189            & 0.0278            & 0.3577            & \textbf{0.8218}   & \textbf{0.0175}   & \textbf{0.0217}   & 0.3621            & 0.9178            & 0.0147            & 0.0177 \\

\bottomrule
\end{tabular}
}
\label{table:category_results}
\end{table*}

In this section, we present and discuss the \var experimental results. 
As for the recommendation results, we evaluate the top-$20$, and top-$50$ recommendation lists since they correspond to $26$ and $48$ fashion items shown on smartphones and desktop navigation on \textit{Amazon.com}, respectively. In the remainder of this section, we may adopt the notation <dataset, VRS, attack, defense> to indicate a specific VAR configuration, where each field in the quadruple may vary depending on the datasets and methods described in Section~\ref{sec:experiment}.

\textit{\textbf{Analysis of the effectiveness of defense mechanisms in protecting  the model from adversarial attacks against IFEs.}}
We start from analyzing the experimental results shown in Table ~\ref{table:image_metrics} (i.e., on attack's $SR$, $FL$) and Table~\ref{table:category_results} (i.e., \chr, \cndcg).

\textbf{Analysis of Attack Success Rate.} We start the \var analysis by exploring the success rate of experimented attacks in fooling the IFE with or without the adversarial robustification techniques. Results showed in Table ~\ref{table:image_metrics} confirm PGD and C \& W as the strongest attacks when applied to lower the classification accuracy of a defense-free CNN classifier. For instance, PGD ($\epsilon = 8$) reaches the $100\%$ of $SR$ for all the studied datasets, while C \& W's $SR$ is always more than $89\%$. As expected, this behavior is different when \var is tested with defense strategies. Under this setting, C \& W emerges as the best offensive solution against defense strategies, as already demonstrated in~\cite{DBLP:conf/sp/Carlini017}. As an example, we observe an average $SR$ reduction in the $SR$ results of $75\%$ for FGSM-methods and $79\%$ for PGD, while it decreases by $43\%$ for C \& W.

Hence, we compare the $SR$ results to the variation of visual-aware recommendations of the products belonging to the perturbed category of images. Table~\ref{table:category_results} presents the results of the proposed \var rank-based evaluation with respect to the origin-target attack scenarios defined in Table~\ref{table:category}.
Quite surprisingly, Table~\ref{table:category_results} shows PGD attacks as extremely more incisive than C \& W in the defense-free settings, i.e., the average value of \chratventi for PGD ($\epsilon = 8$) is $1.2612$, while it is $0.5690$ for C \& W. Conversely, this difference is less observable under defense-activated \var settings, where all the attacks share almost comparable results. These outcomes are in contrast with the $SR$, thus we raise the first contribution: \textit{attack success rate is not directly related to the effects on the recommendation performance. In other words, be powerful enough to lead a classifier in mislabelling an origin product image towards a target class does not explain the effects on the recommendation lists.}

\textbf{Analysis of Features Loss.} Motivated by the previous observations, we investigate the Feature Loss ($FL$) between original and attacked samples whose values are displayed in Table~\ref{table:image_metrics}.
Comparing the results in Table~\ref{table:image_metrics} and Table~\ref{table:category_results} we discover a correlation between the variation of $FL$ and the attack efficacy on VRSs.
For instance, PGD and C \& W results about \chr, and \cndcg, are coherent with the differences observed on the $FL$, i.e., the average value of $FL$ for PGD ($\epsilon = 8$) is $0.1470$ and it is $0.0225$ for C \& W. This association is further confirmed by the less oscillating values of \chr and \cndcg under defense-activated \var settings. 
For instance, <\amazonmen, Traditional,  VBPR> and <\amazonmen, (Adversarial Training, Free Adversarial Training),  VBPR> experiments get a \chratventi standard deviation of $0.3950$ and $0.0260$ respectively, i.e., a difference of more than one order of magnitude. This trend holds true also for $FL$ values, i.e., $0.0441$  and $0.0011$, respectively.
Then, we derive the following contribution: \textit{the modification of VRS is closely linked to the magnitude difference between original and perturbed image features. In short, perturbations leading to larger feature modifications may cause a strong influence on the recommendability of the altered product categories}.

\textbf{Analysis of Category-based Performance.}
After having justified the results in Table~\ref{table:category_results}, we discuss the category-based measures across models and datasets.
Studying the \chr and \cndcg from recommenders point of view, FM appears as the least affected model. For instance <\amazonmen, (Adversarial Training, Free Adversarial Training), FM> and <\amazonwomen, Free Adversarial Training, FM> register reductions even in terms of category measures after all the attacks. In general, the average \chratventi variation between each FM attack-free setting and the most effective attack is only $8.66\%$, i.e., a rather small value compared to other models ($127.02\%$ for VBPR experiments).
\textit{We explain the highly negligible impact of attacks against FM by recalling that the implemented version~\cite{DBLP:conf/recsys/Kula15} is not designed to integrate continuous visual features with a large dimensionality (i.e., 2048) unlike VBPR and AMR. }

As regards VBPR, the PGD strategy reaches the larger variation of \chr and \cndcg in the defense-free setting, confirming the previous considerations made about these metrics with the feature loss. For instance, PGD ($\epsilon = 8 $) leads to a 4 times increase of the \chratventi of the attacked category (i.e., suit), and about a 5 times increase of the \cndcgatventi on the \tradesy dataset. 
Additionally, C \& W is the attack with the best average variation on all the rank-based metrics in the defense scenario. For instance, C \& W attacks make \chratventi and \cndcgatventi grow by an average of $69.48\%$, and $91.01\%$, respectively. These results suggest that \textit{the adversarial robustification strategies have not protected VBPR from the injection of perturbed product images despite they got high performance in protecting the classification task.}

The third tested VRS is AMR. We chose this model since it is the first VRS to integrate adversarial protection by design, so we expected to get a limited variation in traditional performance under attack settings. Surprisingly, results show that AMR is prone to the effects of attacks as much as VBPR. For example, PGD ($\epsilon = 8$) method represents the biggest security threat on the VRS in defense-free settings, while C \& W is the best attack when the IFE is defended. 
Moreover, <Free Adversarial Training, AMR> and <Adversarial Training, AMR> do not provide any protection improvements, notwithstanding the two defense techniques applied on both the IFE and the VRS respectively. For instance, the mild \chratventi improvement seen in <C \& W, Free Adversarial Training, AMR> is higher than the one obtained in <C \& W, Free Adversarial Training, VBPR> (i.e., $77.64\%$ and $69.48\%$ respectively), but the latter did not involve any defense method on the VRS. We conclude that \textit{the combination of the state-of-the-art defense techniques against adversarial perturbations applied on both DNNs and VRSs does not preserve the quality of the recommendation.} 

Conclusively, we compare \chr and \cndcg. The calculated values for \cndcg confirm the trends noticed in \chr. Furthermore, we observe \cndcg shows a relatively higher improvement than the one captured by the \chr in any <dataset, attack, defense, recommender> setting. Here, we draw two final considerations: \textit{(i) \cndcg and \chr are two metrics suitable to evaluate the motivating scenario (i.e., an adversary who wants to push a category of products), and (ii) even though attacks have pushed the products into the top-$K$ lists (see \chr results), an argument could be made that these products get very high positions since \cndcg have increased even more that \chr}.

\textit{\textbf{Evaluation of the effects of adversarial defenses on the IFE in the variation of overall recommendation measures.}}
\begin{table}[!t]
\caption{Overall recommendation performance evaluated on the top-$20$ recommendation lists.}
\scalebox{0.75}{
\begin{tabular}{l  l  l  r r r r r}

\toprule

\multirow{2}{*}{\textbf{Dataset}}          & \multirow{2}{*}{\textbf{VRS}}          & \multirow{2}{*}{\begin{tabular}[r]{@{}c@{}}\textbf{Adversarial}\\\textbf{Attack}\end{tabular}}           & \multicolumn{ 5}{c}{\textbf{Image Feature Extractor}} \\ \cline{ 4- 8}

\multicolumn{ 1}{c}{}          & \multicolumn{ 1}{c}{}          & \multicolumn{ 1}{c}{}          & \multicolumn{1}{c}{$Recall$}          & \multicolumn{1}{c}{$nDCG$}          & \multicolumn{1}{c}{$gini$}          & \multicolumn{1}{c}{$EFD$}          & \multicolumn{1}{c}{$iCov$} \\

\bottomrule
\multicolumn{ 1}{c}{}          & \multicolumn{ 1}{c}{}          & \multicolumn{ 1}{c}{}          & \multicolumn{ 5}{c}{\textbf{Traditional}} \\ \cline{ 2- 8}

\multirow{29}{*}{\begin{tabular}[l]{@{}c@{}}\texttt{Amazon}\\\texttt{Men}\end{tabular}}& 
                            \multirow{3}{*}{FM}          &Original                     & \textbf{0.0027} & \textbf{0.0010} & 0.6025          & \textbf{0.0018} & 7367 \\ 
\multicolumn{ 1}{c}{}&\multicolumn{ 1}{c}{}            &PGD ($\epsilon=8$)             & 0.0024          & 0.0009          & 0.5995          & 0.0016          & 7369 \\ 
\multicolumn{ 1}{c}{}&\multicolumn{ 1}{c}{}            &C \& W                         & 0.0023          & 0.0009          & \textbf{0.6026} & 0.0016          & \textbf{7370}\\ \cline{ 2- 8}
\multicolumn{ 1}{c}{}&\multirow{3}{*}{VBPR}            &Original                       & 0.0127          & \textbf{0.0046} & 0.0519          & \textbf{0.0073} & \textbf{1954} \\ 
\multicolumn{ 1}{c}{}&\multicolumn{ 1}{c}{}            &PGD ($\epsilon=8$)             & 0.0126          & 0.0044          & \textbf{0.0531} & 0.0070          & 1951 \\ 
\multicolumn{ 1}{c}{}&\multicolumn{ 1}{c}{}            &C \& W                         & \textbf{0.0132} & 0.0045          & 0.0519          & 0.0072          & 1925 \\ \cline{ 2- 8}
\multicolumn{ 1}{c}{}&\multirow{3}{*}{AMR}             &Original                       & 0.0330          & 0.0123          & 0.0272          & 0.0171          & 1586 \\ 
\multicolumn{ 1}{c}{}&\multicolumn{ 1}{c}{}            &PGD ($\epsilon=8$)             & 0.0316          & 0.0119          & \textbf{0.0314} & 0.0166          & \textbf{1716} \\ 
\multicolumn{ 1}{c}{}&\multicolumn{ 1}{c}{}            &C \& W                         & \textbf{0.0333} & \textbf{0.0125} & 0.0256          & \textbf{0.0174} & 1488 \\ \cline{ 2- 8}

\multicolumn{ 1}{c}{}&&&\multicolumn{ 5}{c}{\textbf{Adversarial Training}} \\ \cline{ 2- 8}
\multicolumn{ 1}{c}{}&\multirow{3}{*}{FM}              &Original                       & \textbf{0.0034} & \textbf{0.0012} & \textbf{0.6406} & \textbf{0.0021} & \textbf{7370} \\ 
\multicolumn{ 1}{c}{}&\multicolumn{ 1}{c}{}            &PGD ($\epsilon=8$)             & 0.0023          & 0.0009          & 0.5940          & 0.0016          & 7368 \\ 
\multicolumn{ 1}{c}{}&\multicolumn{ 1}{c}{}            &C \& W                         & 0.0023          & 0.0009          & 0.5828          & 0.0016          & 7368 \\ \cline{ 2- 8}
\multicolumn{ 1}{c}{}&\multirow{3}{*}{VBPR}            &Original                       & \textbf{0.0193} & \textbf{0.0076} & 0.0494          & \textbf{0.0112} & 1826 \\ 
\multicolumn{ 1}{c}{}&\multicolumn{ 1}{c}{}            &PGD ($\epsilon=8$)             & 0.0129          & 0.0045          & \textbf{0.0550} & 0.0072          & \textbf{1990}\\ 
\multicolumn{ 1}{c}{}&\multicolumn{ 1}{c}{}            &C \& W                         & 0.0124          & 0.0045          & 0.0543          & 0.0071          & 1956 \\ \cline{ 2- 8}
\multicolumn{ 1}{c}{}&\multirow{3}{*}{AMR}             &Original                       & 0.0312          & 0.0118          & 0.0039          & 0.0151          & 224 \\ 
\multicolumn{ 1}{c}{}&\multicolumn{ 1}{c}{}            &PGD ($\epsilon=8$)             & \textbf{0.0332} & \textbf{0.0125} & \textbf{0.0279} & \textbf{0.0174} & \textbf{1496} \\ 
\multicolumn{ 1}{c}{}&\multicolumn{ 1}{c}{}            &C \& W                         & 0.0327          & 0.0124          & 0.0277          & 0.0172          & 1429 \\ \cline{ 2- 8}

\multicolumn{ 1}{c}{}&          &           & \multicolumn{ 5}{c}{\textbf{Free Adversarial Training}} \\ \cline{ 2- 8}
\multicolumn{ 1}{c}{}&\multirow{3}{*}{FM}              &Original                       & 0.0024          & 0.0009          & \textbf{0.6329} & 0.0017          & \textbf{7371} \\ 
\multicolumn{ 1}{c}{}&\multicolumn{ 1}{c}{}            &PGD ($\epsilon=8$)             & \textbf{0.0025} & \textbf{0.0010} & 0.6016          & \textbf{0.0018} & 7369 \\ 
\multicolumn{ 1}{c}{}&\multicolumn{ 1}{c}{}            &C \& W                         & 0.0021          & 0.0008          & 0.5991          & 0.0015          & 7366 \\ \cline{ 2- 8}
\multicolumn{ 1}{c}{}&\multirow{3}{*}{VBPR}            &Original                       & \textbf{0.0236} & \textbf{0.0089} & 0.0325          & \textbf{0.0125} & 1184 \\ 
\multicolumn{ 1}{c}{}&\multicolumn{ 1}{c}{}            &PGD ($\epsilon=8$)             & 0.0129          & 0.0046          & 0.0528          & 0.0073          & 1900 \\ 
\multicolumn{ 1}{c}{}&\multicolumn{ 1}{c}{}            &C \& W                         & 0.0123          & 0.0044          & \textbf{0.0555} & 0.0070          & \textbf{1992} \\ \cline{ 2- 8}
\multicolumn{ 1}{c}{}&\multirow{3}{*}{AMR}             &Original                       & 0.0317          & 0.0118          & 0.0067          & 0.0154          & 402 \\ 
\multicolumn{ 1}{c}{}&\multicolumn{ 1}{c}{}            &PGD ($\epsilon=8$)             & \textbf{0.0331}          & \textbf{0.0123}          & \textbf{0.0271}          & \textbf{0.0171}          & 1420 \\ 
\multicolumn{ 1}{c}{}&\multicolumn{ 1}{c}{}            &C \& W                         & 0.0325          & \textbf{0.0123}          & 0.0265          & \textbf{0.0171}          & \textbf{1425} \\ 

\bottomrule

\end{tabular}
}
\label{table:overall_accuracy}
\end{table}

Table~\ref{table:overall_accuracy} shows the accuracy and beyond accuracy results on the \amazonmen dataset for the two most powerful attacks previously discussed. The intuition behind this evaluation is to understand whether the application of adversarial defenses ---adopted to alleviate attacks' influence--- may generate a drastic variation of the overall recommendation performance.
In Table~\ref{table:overall_accuracy}, we bolded the best results for each <defense, VRS> experiment. Surprisingly, we see that \textit{the application of powerful attacks has not tragically worsened the accuracy and beyond accuracy performance. On the contrary, some measures have significantly improved as a consequence of the attack}. For instance, AMR has its best performance under all combinations attacks combinations independently from the application of a defense mechanism. On the other side, it is worth noticing the item coverage generally gets better under attack settings. We explain this effect by the fact that, since adversarial images can be considered a source of randomness, they might help the recommendation of products from the long tail (e.g., products similar to the attacked ones may get benefit from the attack process). 

Furthermore, Table~\ref{table:overall_accuracy} shows that VBPR and FM have their best accuracy performance in defense-activated settings. For instance, <\amazonmen, Original, Defense-free, VBPR> has \textit{Recall} and \textit{nDCG} equal to 0.0127 and 0.0046, while in the defense-activated settings both metrics are quite doubled, i.e., 0.0193 and 0.0076 in Adversarial Training and 0.0236 and 0.0089 in Free Adversarial Training. The confirmation of this trend in attack scenarios raises the following conclusion: \textit{the application of defense mechanisms on the IFE is a system design possibility that preserves, or even improves, the overall performance of a VRS, while does not guarantee the protection from altering the recommendability of the category of products.}

\section{Conclusion and Future Work}\label{sec:conclusion}
We have presented an evaluation framework, i.e., Visual Adversarial Recommendation (\var), to explore the efficacy of adversarial robustification mechanisms against several state-of-the-art adversarial attacks and to investigate the impact of perturbed product images on visually-aware recommendations.
Experimental results have shown that defense mechanisms do not guarantee the protections of recommenders against attacks also in the case of low-success rate attacks. Interestingly, we have found that the effectiveness of attacks in altering the recommenders is more related to high feature losses than high success rates. This finding raises interesting opportunities to develop novel recommender models along with defense strategies. Finally, we have verified that overall recommendation performance has not worsened under the experimented threat model, and (surprisingly) defended IFEs may even improve in non-attack settings. This further opens future directions in finding the reasons behind this behavior, getting the most benefits from it.

\newpage
\bibliographystyle{ACM-Reference-Format}
\bibliography{bibliography}


\begin{thebibliography}{41}


\ifx \showCODEN    \undefined \def \showCODEN     #1{\unskip}     \fi
\ifx \showDOI      \undefined \def \showDOI       #1{#1}\fi
\ifx \showISBNx    \undefined \def \showISBNx     #1{\unskip}     \fi
\ifx \showISBNxiii \undefined \def \showISBNxiii  #1{\unskip}     \fi
\ifx \showISSN     \undefined \def \showISSN      #1{\unskip}     \fi
\ifx \showLCCN     \undefined \def \showLCCN      #1{\unskip}     \fi
\ifx \shownote     \undefined \def \shownote      #1{#1}          \fi
\ifx \showarticletitle \undefined \def \showarticletitle #1{#1}   \fi
\ifx \showURL      \undefined \def \showURL       {\relax}        \fi
\providecommand\bibfield[2]{#2}
\providecommand\bibinfo[2]{#2}
\providecommand\natexlab[1]{#1}
\providecommand\showeprint[2][]{arXiv:#2}

\bibitem[\protect\citeauthoryear{Athalye, Carlini, and Wagner}{Athalye
  et~al\mbox{.}}{2018}]%
        {DBLP:conf/icml/AthalyeC018}
\bibfield{author}{\bibinfo{person}{Anish Athalye}, \bibinfo{person}{Nicholas
  Carlini}, {and} \bibinfo{person}{David~A. Wagner}.}
  \bibinfo{year}{2018}\natexlab{}.
\newblock \showarticletitle{Obfuscated Gradients Give a False Sense of
  Security: Circumventing Defenses to Adversarial Examples}. In
  \bibinfo{booktitle}{\emph{{ICML} 2018}}.
\newblock


\bibitem[\protect\citeauthoryear{Bhaumik, Williams, Mobasher, and
  Burke}{Bhaumik et~al\mbox{.}}{2006}]%
        {bhaumik2006securing}
\bibfield{author}{\bibinfo{person}{Runa Bhaumik}, \bibinfo{person}{Chad
  Williams}, \bibinfo{person}{Bamshad Mobasher}, {and} \bibinfo{person}{Robin
  Burke}.} \bibinfo{year}{2006}\natexlab{}.
\newblock \showarticletitle{Securing collaborative filtering against malicious
  attacks through anomaly detection}. In \bibinfo{booktitle}{\emph{{ITWP}
  2006}}.
\newblock


\bibitem[\protect\citeauthoryear{Biggio, Corona, Maiorca, Nelson, Srndic,
  Laskov, Giacinto, and Roli}{Biggio et~al\mbox{.}}{2013}]%
        {DBLP:conf/pkdd/BiggioCMNSLGR13}
\bibfield{author}{\bibinfo{person}{Battista Biggio}, \bibinfo{person}{Igino
  Corona}, \bibinfo{person}{Davide Maiorca}, \bibinfo{person}{Blaine Nelson},
  \bibinfo{person}{Nedim Srndic}, \bibinfo{person}{Pavel Laskov},
  \bibinfo{person}{Giorgio Giacinto}, {and} \bibinfo{person}{Fabio Roli}.}
  \bibinfo{year}{2013}\natexlab{}.
\newblock \showarticletitle{Evasion Attacks against Machine Learning at Test
  Time}. In \bibinfo{booktitle}{\emph{{ECML}-{PKDD} 2013}}.
\newblock


\bibitem[\protect\citeauthoryear{Carlini, Athalye, Papernot, Brendel, Rauber,
  Tsipras, Goodfellow, Madry, and Kurakin}{Carlini et~al\mbox{.}}{2019}]%
        {DBLP:journals/corr/abs-1902-06705}
\bibfield{author}{\bibinfo{person}{Nicholas Carlini}, \bibinfo{person}{Anish
  Athalye}, \bibinfo{person}{Nicolas Papernot}, \bibinfo{person}{Wieland
  Brendel}, \bibinfo{person}{Jonas Rauber}, \bibinfo{person}{Dimitris Tsipras},
  \bibinfo{person}{Ian~J. Goodfellow}, \bibinfo{person}{Aleksander Madry},
  {and} \bibinfo{person}{Alexey Kurakin}.} \bibinfo{year}{2019}\natexlab{}.
\newblock \showarticletitle{On Evaluating Adversarial Robustness}.
\newblock \bibinfo{journal}{\emph{{CoRR} 2019}} (\bibinfo{year}{2019}).
\newblock


\bibitem[\protect\citeauthoryear{Carlini and Wagner}{Carlini and
  Wagner}{2016}]%
        {DBLP:journals/corr/CarliniW16}
\bibfield{author}{\bibinfo{person}{Nicholas Carlini} {and}
  \bibinfo{person}{David~A. Wagner}.} \bibinfo{year}{2016}\natexlab{}.
\newblock \showarticletitle{Defensive Distillation is Not Robust to Adversarial
  Examples}.
\newblock \bibinfo{journal}{\emph{{CoRR} 2016}} (\bibinfo{year}{2016}).
\newblock


\bibitem[\protect\citeauthoryear{Carlini and Wagner}{Carlini and
  Wagner}{2017a}]%
        {DBLP:conf/ccs/Carlini017}
\bibfield{author}{\bibinfo{person}{Nicholas Carlini} {and}
  \bibinfo{person}{David~A. Wagner}.} \bibinfo{year}{2017}\natexlab{a}.
\newblock \showarticletitle{Adversarial Examples Are Not Easily Detected:
  Bypassing Ten Detection Methods}. In \bibinfo{booktitle}{\emph{{AISec@CCS}
  2017}}.
\newblock


\bibitem[\protect\citeauthoryear{Carlini and Wagner}{Carlini and
  Wagner}{2017b}]%
        {DBLP:conf/sp/Carlini017}
\bibfield{author}{\bibinfo{person}{Nicholas Carlini} {and}
  \bibinfo{person}{David~A. Wagner}.} \bibinfo{year}{2017}\natexlab{b}.
\newblock \showarticletitle{Towards Evaluating the Robustness of Neural
  Networks}. In \bibinfo{booktitle}{\emph{{SP} 2017}}.
\newblock


\bibitem[\protect\citeauthoryear{Chu and Tsai}{Chu and Tsai}{2017}]%
        {DBLP:journals/www/ChuT17}
\bibfield{author}{\bibinfo{person}{Wei{-}Ta Chu} {and}
  \bibinfo{person}{Ya{-}Lun Tsai}.} \bibinfo{year}{2017}\natexlab{}.
\newblock \showarticletitle{A hybrid recommendation system considering visual
  information for predicting favorite restaurants}.
\newblock \bibinfo{journal}{\emph{{WWW} 2017}} (\bibinfo{year}{2017}).
\newblock


\bibitem[\protect\citeauthoryear{Deldjoo, Noia, and Merra}{Deldjoo
  et~al\mbox{.}}{2020}]%
        {DBLP:conf/wsdm/DeldjooNM20}
\bibfield{author}{\bibinfo{person}{Yashar Deldjoo}, \bibinfo{person}{Tommaso~Di
  Noia}, {and} \bibinfo{person}{Felice~Antonio Merra}.}
  \bibinfo{year}{2020}\natexlab{}.
\newblock \showarticletitle{Adversarial Machine Learning Recommender Systems
  (AML-RecSys)}. In \bibinfo{booktitle}{\emph{{WSDM} 2020}}.
\newblock


\bibitem[\protect\citeauthoryear{{Di Noia}, Malitesta, and Merra}{{Di Noia}
  et~al\mbox{.}}{2020}]%
        {DSML20}
\bibfield{author}{\bibinfo{person}{Tommaso {Di Noia}}, \bibinfo{person}{Daniele
  Malitesta}, {and} \bibinfo{person}{Felice~Antonio Merra}.}
  \bibinfo{year}{2020}\natexlab{}.
\newblock \showarticletitle{TAaMR: Targeted Adversarial Attack against
  Multimedia Recommender Systems}. In \bibinfo{booktitle}{\emph{DSN–DSML
  2020}}.
\newblock


\bibitem[\protect\citeauthoryear{Entezari, Al{-}Sayouri, Darvishzadeh, and
  Papalexakis}{Entezari et~al\mbox{.}}{2020}]%
        {DBLP:conf/wsdm/EntezariADP20}
\bibfield{author}{\bibinfo{person}{Negin Entezari}, \bibinfo{person}{Saba~A.
  Al{-}Sayouri}, \bibinfo{person}{Amirali Darvishzadeh}, {and}
  \bibinfo{person}{Evangelos~E. Papalexakis}.} \bibinfo{year}{2020}\natexlab{}.
\newblock \showarticletitle{All You Need Is Low (Rank): Defending Against
  Adversarial Attacks on Graphs}. In \bibinfo{booktitle}{\emph{{WSDM} 2020}}.
\newblock


\bibitem[\protect\citeauthoryear{Goodfellow and Bengio}{Goodfellow and
  Bengio}{2017}]%
        {DBLP:conf/iclr/KurakinGB17a}
\bibfield{author}{\bibinfo{person}{Alexey Kurakand Ian~J. Goodfellow} {and}
  \bibinfo{person}{Samy Bengio}.} \bibinfo{year}{2017}\natexlab{}.
\newblock \showarticletitle{Adversarial examples the physical world}. In
  \bibinfo{booktitle}{\emph{{ICLR} 2017}}.
\newblock


\bibitem[\protect\citeauthoryear{Goodfellow, Shlens, and Szegedy}{Goodfellow
  et~al\mbox{.}}{2015}]%
        {DBLP:journals/corr/GoodfellowSS14}
\bibfield{author}{\bibinfo{person}{Ian~J. Goodfellow},
  \bibinfo{person}{Jonathon Shlens}, {and} \bibinfo{person}{Christian
  Szegedy}.} \bibinfo{year}{2015}\natexlab{}.
\newblock \showarticletitle{Explaining and Harnessing Adversarial Examples}. In
  \bibinfo{booktitle}{\emph{{ICLR} 2015}}.
\newblock


\bibitem[\protect\citeauthoryear{Grauman}{Grauman}{2020}]%
        {DBLP:conf/wsdm/Grauman20}
\bibfield{author}{\bibinfo{person}{Kristen Grauman}.}
  \bibinfo{year}{2020}\natexlab{}.
\newblock \showarticletitle{Computer Vision for Fashion: From Individual
  Recommendations to World-wide Trends}. In \bibinfo{booktitle}{\emph{{WSDM}
  2020}}.
\newblock


\bibitem[\protect\citeauthoryear{Guo, Rana, Ciss{\'{e}}, and van~der
  Maaten}{Guo et~al\mbox{.}}{2018}]%
        {DBLP:conf/iclr/GuoRCM18}
\bibfield{author}{\bibinfo{person}{Chuan Guo}, \bibinfo{person}{Mayank Rana},
  \bibinfo{person}{Moustapha Ciss{\'{e}}}, {and} \bibinfo{person}{Laurens
  van~der Maaten}.} \bibinfo{year}{2018}\natexlab{}.
\newblock \showarticletitle{Countering Adversarial Images using Input
  Transformations}. In \bibinfo{booktitle}{\emph{{ICLR} 2018}}.
\newblock


\bibitem[\protect\citeauthoryear{He, Zhang, Ren, and Sun}{He
  et~al\mbox{.}}{2016}]%
        {DBLP:conf/cvpr/HeZRS16}
\bibfield{author}{\bibinfo{person}{Kaiming He}, \bibinfo{person}{Xiangyu
  Zhang}, \bibinfo{person}{Shaoqing Ren}, {and} \bibinfo{person}{Jian Sun}.}
  \bibinfo{year}{2016}\natexlab{}.
\newblock \showarticletitle{Deep Residual Learning for Image Recognition}. In
  \bibinfo{booktitle}{\emph{{CVPR} 2016}}.
\newblock


\bibitem[\protect\citeauthoryear{He and McAuley}{He and McAuley}{2016a}]%
        {DBLP:conf/www/HeM16}
\bibfield{author}{\bibinfo{person}{Ruining He} {and} \bibinfo{person}{Julian~J.
  McAuley}.} \bibinfo{year}{2016}\natexlab{a}.
\newblock \showarticletitle{Ups and Downs: Modeling the Visual Evolution of
  Fashion Trends with One-Class Collaborative Filtering}. In
  \bibinfo{booktitle}{\emph{{WWW} 2016}}.
\newblock


\bibitem[\protect\citeauthoryear{He and McAuley}{He and McAuley}{2016b}]%
        {DBLP:conf/aaai/HeM16}
\bibfield{author}{\bibinfo{person}{Ruining He} {and} \bibinfo{person}{Julian~J.
  McAuley}.} \bibinfo{year}{2016}\natexlab{b}.
\newblock \showarticletitle{{VBPR:} Visual Bayesian Personalized Ranking from
  Implicit Feedback}. In \bibinfo{booktitle}{\emph{{AAAI} 2016}}.
\newblock


\bibitem[\protect\citeauthoryear{He, He, Du, and Chua}{He
  et~al\mbox{.}}{2018}]%
        {DBLP:conf/sigir/0001HDC18}
\bibfield{author}{\bibinfo{person}{Xiangnan He}, \bibinfo{person}{Zhankui He},
  \bibinfo{person}{Xiaoyu Du}, {and} \bibinfo{person}{Tat{-}Seng Chua}.}
  \bibinfo{year}{2018}\natexlab{}.
\newblock \showarticletitle{Adversarial Personalized Ranking for
  Recommendation}. In \bibinfo{booktitle}{\emph{{SIGIR} 2018}}.
\newblock


\bibitem[\protect\citeauthoryear{{Hinton}, {Deng}, {Yu}, {Dahl}, {Mohamed},
  {Jaitly}, {Senior}, {Vanhoucke}, {Nguyen}, {Sainath}, and
  {Kingsbury}}{{Hinton} et~al\mbox{.}}{2012}]%
        {speech}
\bibfield{author}{\bibinfo{person}{G. {Hinton}}, \bibinfo{person}{L. {Deng}},
  \bibinfo{person}{D. {Yu}}, \bibinfo{person}{G.~E. {Dahl}},
  \bibinfo{person}{A. {Mohamed}}, \bibinfo{person}{N. {Jaitly}},
  \bibinfo{person}{A. {Senior}}, \bibinfo{person}{V. {Vanhoucke}},
  \bibinfo{person}{P. {Nguyen}}, \bibinfo{person}{T.~N. {Sainath}}, {and}
  \bibinfo{person}{B. {Kingsbury}}.} \bibinfo{year}{2012}\natexlab{}.
\newblock \showarticletitle{Deep Neural Networks for Acoustic Modeling Speech
  Recognition: The Shared Views of Four Research Groups}.
\newblock \bibinfo{journal}{\emph{IEEE Signal Processing Magazine}}
  (\bibinfo{year}{2012}).
\newblock


\bibitem[\protect\citeauthoryear{Kang, Fang, Wang, and McAuley}{Kang
  et~al\mbox{.}}{[n.d.]}]%
        {DBLP:conf/icdm/KangFWM17}
\bibfield{author}{\bibinfo{person}{Wang{-}Cheng Kang}, \bibinfo{person}{Chen
  Fang}, \bibinfo{person}{Zhaowen Wang}, {and} \bibinfo{person}{Julian~J.
  McAuley}.} \bibinfo{year}{[n.d.]}\natexlab{}.
\newblock \showarticletitle{Visually-Aware Fashion Recommendation and Design
  with Generative Image Models}. In \bibinfo{booktitle}{\emph{{ICDM} 2017}}.
\newblock


\bibitem[\protect\citeauthoryear{Kordan and Kotov}{Kordan and Kotov}{2018}]%
        {DBLP:conf/wsdm/KordanK18}
\bibfield{author}{\bibinfo{person}{Saeid~Balaneshin Kordan} {and}
  \bibinfo{person}{Alexander Kotov}.} \bibinfo{year}{2018}\natexlab{}.
\newblock \showarticletitle{Deep Neural Architecture for Multi-Modal Retrieval
  based on Joint Embedding Space for Text and Images}. In
  \bibinfo{booktitle}{\emph{{WSDM} 2018}}.
\newblock


\bibitem[\protect\citeauthoryear{Kula}{Kula}{2015}]%
        {DBLP:conf/recsys/Kula15}
\bibfield{author}{\bibinfo{person}{Maciej Kula}.}
  \bibinfo{year}{2015}\natexlab{}.
\newblock \showarticletitle{Metadata Embeddings for User and Item Cold-start
  Recommendations}. In \bibinfo{booktitle}{\emph{{CBRecSys@RecSys} 2015}}.
\newblock


\bibitem[\protect\citeauthoryear{Lam and Riedl}{Lam and Riedl}{2004}]%
        {DBLP:conf/www/LamR04}
\bibfield{author}{\bibinfo{person}{Shyong~K. Lam} {and} \bibinfo{person}{John
  Riedl}.} \bibinfo{year}{2004}\natexlab{}.
\newblock \showarticletitle{Shilling recommender systems for fun and profit}.
  In \bibinfo{booktitle}{\emph{{WWW} 2004}}.
\newblock


\bibitem[\protect\citeauthoryear{Madry, Makelov, Schmidt, Tsipras, and
  Vladu}{Madry et~al\mbox{.}}{2018}]%
        {DBLP:conf/iclr/MadryMSTV18}
\bibfield{author}{\bibinfo{person}{Aleksander Madry},
  \bibinfo{person}{Aleksandar Makelov}, \bibinfo{person}{Ludwig Schmidt},
  \bibinfo{person}{Dimitris Tsipras}, {and} \bibinfo{person}{Adrian Vladu}.}
  \bibinfo{year}{2018}\natexlab{}.
\newblock \showarticletitle{Towards Deep Learning Models Resistant to
  Adversarial Attacks}. In \bibinfo{booktitle}{\emph{{ICLR} 2018}}.
\newblock


\bibitem[\protect\citeauthoryear{Niu, Caverlee, and Lu}{Niu
  et~al\mbox{.}}{2018}]%
        {DBLP:conf/wsdm/NiuCL18}
\bibfield{author}{\bibinfo{person}{Wei Niu}, \bibinfo{person}{James Caverlee},
  {and} \bibinfo{person}{Haokai Lu}.} \bibinfo{year}{2018}\natexlab{}.
\newblock \showarticletitle{Neural Personalized Ranking for Image
  Recommendation}. In \bibinfo{booktitle}{\emph{{WSDM} 2018}}.
\newblock


\bibitem[\protect\citeauthoryear{O'Mahony, Hurley, Kushmerick, and
  Silvestre}{O'Mahony et~al\mbox{.}}{2004}]%
        {DBLP:journals/toit/OMahonyHKS04}
\bibfield{author}{\bibinfo{person}{Michael~P. O'Mahony},
  \bibinfo{person}{Neil~J. Hurley}, \bibinfo{person}{Nicholas Kushmerick},
  {and} \bibinfo{person}{Guenole C.~M. Silvestre}.}
  \bibinfo{year}{2004}\natexlab{}.
\newblock \showarticletitle{Collaborative recommendation: {A} robustness
  analysis}.
\newblock \bibinfo{journal}{\emph{{ACM} Trans. Internet Techn.}}
  (\bibinfo{year}{2004}).
\newblock


\bibitem[\protect\citeauthoryear{Papernot, Faghri, Carlini, Goodfellow,
  Feinman, Xie, Sharma, Brown, Roy, Matyasko, Behzadan, Hambardzumyan, Zhang,
  Yi-LJuang, Li, Sheatsley, Garg, Uesato, Gierke, Dong, Berthelot, Hendricks,
  Rauber, and Long}{Papernot et~al\mbox{.}}{2018}]%
        {papernot2018cleverhans}
\bibfield{author}{\bibinfo{person}{Nicolas Papernot}, \bibinfo{person}{Fartash
  Faghri}, \bibinfo{person}{Nicholas Carlini}, \bibinfo{person}{Ian
  Goodfellow}, \bibinfo{person}{Reuben Feinman}, \bibinfo{person}{Alexey
  Kurakand~Cihang Xie}, \bibinfo{person}{Yash Sharma}, \bibinfo{person}{Tom
  Brown}, \bibinfo{person}{Aurko Roy}, \bibinfo{person}{Alexander Matyasko},
  \bibinfo{person}{Vahid Behzadan}, \bibinfo{person}{Karen Hambardzumyan},
  \bibinfo{person}{Zhishuai Zhang}, \bibinfo{person}{Yi-LJuang},
  \bibinfo{person}{Zhi Li}, \bibinfo{person}{Ryan Sheatsley},
  \bibinfo{person}{Abhibhav Garg}, \bibinfo{person}{Jonathan Uesato},
  \bibinfo{person}{Willi Gierke}, \bibinfo{person}{Yinpeng Dong},
  \bibinfo{person}{David Berthelot}, \bibinfo{person}{Paul Hendricks},
  \bibinfo{person}{Jonas Rauber}, {and} \bibinfo{person}{Rujun Long}.}
  \bibinfo{year}{2018}\natexlab{}.
\newblock \showarticletitle{Technical Report on the CleverHans v2.1.0
  Adversarial Examples Library}.
\newblock \bibinfo{journal}{\emph{{Corr} 2018}} (\bibinfo{year}{2018}).
\newblock


\bibitem[\protect\citeauthoryear{Papernot, McDaniel, Wu, Jha, and
  Swami}{Papernot et~al\mbox{.}}{2016}]%
        {DBLP:conf/sp/PapernotM0JS16}
\bibfield{author}{\bibinfo{person}{Nicolas Papernot},
  \bibinfo{person}{Patrick~D. McDaniel}, \bibinfo{person}{Xi Wu},
  \bibinfo{person}{Somesh Jha}, {and} \bibinfo{person}{Ananthram Swami}.}
  \bibinfo{year}{2016}\natexlab{}.
\newblock \showarticletitle{Distillation as a Defense to Adversarial
  Perturbations Against Deep Neural Networks}. In
  \bibinfo{booktitle}{\emph{{SP} 2016}}.
\newblock


\bibitem[\protect\citeauthoryear{Ren, He, Girshick, and Sun}{Ren
  et~al\mbox{.}}{2015}]%
        {DBLP:conf/nips/RenHGS15}
\bibfield{author}{\bibinfo{person}{Shaoqing Ren}, \bibinfo{person}{Kaiming He},
  \bibinfo{person}{Ross~B. Girshick}, {and} \bibinfo{person}{Jian Sun}.}
  \bibinfo{year}{2015}\natexlab{}.
\newblock \showarticletitle{Faster {R-CNN:} Towards Real-Time Object Detection
  with Region Proposal Networks}. In \bibinfo{booktitle}{\emph{{NeurIPS}
  2015}}.
\newblock


\bibitem[\protect\citeauthoryear{Rendle}{Rendle}{2010}]%
        {DBLP:conf/icdm/Rendle10}
\bibfield{author}{\bibinfo{person}{Steffen Rendle}.}
  \bibinfo{year}{2010}\natexlab{}.
\newblock \showarticletitle{Factorization Machines}. In
  \bibinfo{booktitle}{\emph{{ICDM} 2010}}.
\newblock


\bibitem[\protect\citeauthoryear{Rendle, Freudenthaler, Gantner, and
  Schmidt{-}Thieme}{Rendle et~al\mbox{.}}{209}]%
        {DBLP:conf/uai/RendleFGS09}
\bibfield{author}{\bibinfo{person}{Steffen Rendle}, \bibinfo{person}{Christoph
  Freudenthaler}, \bibinfo{person}{Zeno Gantner}, {and} \bibinfo{person}{Lars
  Schmidt{-}Thieme}.} \bibinfo{year}{209}\natexlab{}.
\newblock \showarticletitle{{BPR:} Bayesian Personalized Ranking from Implicit
  Feedback}. In \bibinfo{booktitle}{\emph{{UAI} 2009}}.
\newblock


\bibitem[\protect\citeauthoryear{Ricci, Rokach, and Shapira}{Ricci
  et~al\mbox{.}}{2015}]%
        {DBLP:reference/sp/2015rsh}
\bibfield{editor}{\bibinfo{person}{Francesco Ricci}, \bibinfo{person}{Lior
  Rokach}, {and} \bibinfo{person}{Bracha Shapira}} (Eds.).
  \bibinfo{year}{2015}\natexlab{}.
\newblock \bibinfo{booktitle}{\emph{Recommender Systems Handbook}}.
\newblock \bibinfo{publisher}{Springer}.
\newblock


\bibitem[\protect\citeauthoryear{Shafahi, Najibi, AmGhiasi, Xu, Dickerson,
  Studer, Davis, GavTaylor, and Goldstein}{Shafahi et~al\mbox{.}}{2019}]%
        {DBLP:conf/nips/ShafahiNG0DSDTG19}
\bibfield{author}{\bibinfo{person}{Ali Shafahi}, \bibinfo{person}{Mahyar
  Najibi}, \bibinfo{person}{AmGhiasi}, \bibinfo{person}{Zheng Xu},
  \bibinfo{person}{John~P. Dickerson}, \bibinfo{person}{Christoph Studer},
  \bibinfo{person}{Larry~S. Davis}, \bibinfo{person}{GavTaylor}, {and}
  \bibinfo{person}{Tom Goldstein}.} \bibinfo{year}{2019}\natexlab{}.
\newblock \showarticletitle{Adversarial training for free!}. In
  \bibinfo{booktitle}{\emph{{NeurIPS} 2019}}.
\newblock


\bibitem[\protect\citeauthoryear{Szegedy, Zaremba, Sutskever, Bruna, Erhan,
  Goodfellow, and Fergus}{Szegedy et~al\mbox{.}}{2014}]%
        {DBLP:journals/corr/SzegedyZSBEGF13}
\bibfield{author}{\bibinfo{person}{Christian Szegedy},
  \bibinfo{person}{Wojciech Zaremba}, \bibinfo{person}{Ilya Sutskever},
  \bibinfo{person}{Joan Bruna}, \bibinfo{person}{Dumitru Erhan},
  \bibinfo{person}{Ian~J. Goodfellow}, {and} \bibinfo{person}{Rob Fergus}.}
  \bibinfo{year}{2014}\natexlab{}.
\newblock \showarticletitle{Intriguing properties of neural networks}. In
  \bibinfo{booktitle}{\emph{{ICLR} 2014}}.
\newblock


\bibitem[\protect\citeauthoryear{{Tang}, {Du}, {He}, {Yuan}, {Tian}, and
  {Chua}}{{Tang} et~al\mbox{.}}{2019}]%
        {8618394}
\bibfield{author}{\bibinfo{person}{J. {Tang}}, \bibinfo{person}{X. {Du}},
  \bibinfo{person}{X. {He}}, \bibinfo{person}{F. {Yuan}}, \bibinfo{person}{Q.
  {Tian}}, {and} \bibinfo{person}{T. {Chua}}.} \bibinfo{year}{2019}\natexlab{}.
\newblock \showarticletitle{Adversarial Training Towards Robust Multimedia
  Recommender System}.
\newblock \bibinfo{journal}{\emph{{TKDE} 2019}} (\bibinfo{year}{2019}).
\newblock


\bibitem[\protect\citeauthoryear{Tang, Li, Sun, Yao, Mitra, and Wang}{Tang
  et~al\mbox{.}}{2020}]%
        {DBLP:conf/wsdm/TangLSYMW20}
\bibfield{author}{\bibinfo{person}{Xianfeng Tang}, \bibinfo{person}{Yandong
  Li}, \bibinfo{person}{Yiwei Sun}, \bibinfo{person}{Huaxiu Yao},
  \bibinfo{person}{Prasenjit Mitra}, {and} \bibinfo{person}{Suhang Wang}.}
  \bibinfo{year}{2020}\natexlab{}.
\newblock \showarticletitle{Transferring Robustness for Graph Neural Network
  Against Poisoning Attacks}. In \bibinfo{booktitle}{\emph{{WSDM} 2020}}.
\newblock


\bibitem[\protect\citeauthoryear{Wang, YilWang, Tang, Shu, Ranganath, and
  Liu}{Wang et~al\mbox{.}}{2017}]%
        {DBLP:conf/www/WangWTSRL17}
\bibfield{author}{\bibinfo{person}{Suhang Wang}, \bibinfo{person}{YilWang},
  \bibinfo{person}{Jiliang Tang}, \bibinfo{person}{Kai Shu},
  \bibinfo{person}{Suhas Ranganath}, {and} \bibinfo{person}{Huan Liu}.}
  \bibinfo{year}{2017}\natexlab{}.
\newblock \showarticletitle{What Your Images Reveal: Exploiting Visual Contents
  for Point-of-Interest Recommendation}. In \bibinfo{booktitle}{\emph{{WWW}
  2017}}.
\newblock


\bibitem[\protect\citeauthoryear{Wu, Liu, Zhang, Wu, Zhang, and Ma}{Wu
  et~al\mbox{.}}{2019}]%
        {DBLP:conf/wsdm/WuLZWZM19}
\bibfield{author}{\bibinfo{person}{Zhijing Wu}, \bibinfo{person}{Yiqun Liu},
  \bibinfo{person}{Qianfan Zhang}, \bibinfo{person}{Kailu Wu},
  \bibinfo{person}{Min Zhang}, {and} \bibinfo{person}{Shaoping Ma}.}
  \bibinfo{year}{2019}\natexlab{}.
\newblock \showarticletitle{The Influence of Image Search Intents on User
  Behavior and Satisfaction}. In \bibinfo{booktitle}{\emph{{WSDM} 2019}}.
\newblock


\bibitem[\protect\citeauthoryear{Yuan, Lu, Wang, and Xue}{Yuan
  et~al\mbox{.}}{2014}]%
        {DBLP:conf/sigcomm/YuanLWX14}
\bibfield{author}{\bibinfo{person}{Zhenlong Yuan}, \bibinfo{person}{Yongqiang
  Lu}, \bibinfo{person}{Zhaoguo Wang}, {and} \bibinfo{person}{Yibo Xue}.}
  \bibinfo{year}{2014}\natexlab{}.
\newblock \showarticletitle{Droid-Sec: deep learning android malware
  detection}. In \bibinfo{booktitle}{\emph{{SIGCOMM} 2014}}.
\newblock


\bibitem[\protect\citeauthoryear{YZhang and Caverlee}{YZhang and
  Caverlee}{2019}]%
        {DBLP:conf/cikm/ZhangC19}
\bibfield{author}{\bibinfo{person}{YZhang} {and} \bibinfo{person}{James
  Caverlee}.} \bibinfo{year}{2019}\natexlab{}.
\newblock \showarticletitle{Instagrammers, Fashionistas, and Me: Recurrent
  Fashion Recommendation with Implicit Visual Influence}. In
  \bibinfo{booktitle}{\emph{{CIKM} 2019}}.
\newblock


\end{thebibliography}

\end{document}